\documentstyle[twoside,fleqn,epsf,espcrc2]{article}

\newcommand{\csw}{$C_{\rm SW}$}
\newcommand{\dfrac}{\displaystyle\frac}
\newcommand{\mPS}{m_{\rm PS}}
\newcommand{\mPSv}{m_{\rm PS,val}}
\newcommand{\mPSs}{m_{\rm PS,s}}
\newcommand{\mV}{m_{\rm V}}
\newcommand{\mN}{m_{\rm N}}
\newcommand{\mH}{m_{\rm H}}
\newcommand{\nn}{\nonumber}
\title{Unquenched QCD simulation results}

\author{Sinya Aoki\address{Institute of Physics, 
        University of Tsukuba, \\ 
        Tsukuba, Ibaraki 305-8571, JAPAN}%
        \thanks{saoki@het.ph.tsukuba.ac.jp}
         }
       
\begin{document}
\begin{abstract}
The recent progress on unquenched QCD simulations is critically reviewed.
After some discussions on problems and subtleties in unquenched simulations,
hadron spectra obtained from both quenched and unquenched simulations
are compared among various gauge and quark actions.
It is found that the Edinburgh plot does not agree in the continuum limit 
between Wilson and KS quark actions even in quenched QCD.
Dynamical quark effects on hadron spectra, in particular, on meson masses
are then presented for Wilson-type quark actions.
Finally dynamical quark effects on other quantities such as
the topological susceptibility and the flavor-singlet meson mass are
discussed.
\vspace{1pc}
\end{abstract}

\maketitle

\section{Introduction}
An ultimate goal of lattice QCD simulations is to numerically solve
the dynamics of the strong interaction.
Due to lack of enough computational resources, however,
a large part of simulations so far have been performed within the quenched 
approximation.
Recent results from high statistical simulations\cite{CPPACS_quench}
identified errors associated with the quenched approximation:
predicted hadron masses in the continuum limit, in particular
those containing strange quarks, differ from corresponding experimental values 
as large as 10\%.
The effect of the quark determinant must remove these differences.

In this talk I review recent progress in unquenched
QCD simulations.
I will survey the quality of current unquenched simulations
and to what extent dynamical quark effects can be seen,
in the light of the precise results from the quenched 
simulation\cite{CPPACS_quench}.

I first discuss several problems in unquenched simulations such as
the auto-correlation and the finite size effect, which 
slowed down a lot unquenched simulations in the past. 
Although these problems
still exist, they are not a main obstruction anymore.
A brief consideration on the scale determination and a 
theoretical discussion on the problem of
changing the quark mass in unquenched QCD then follow.

Using a large part of this report hadron spectra are analyzed
for KS and Wilson quark actions in parallel, to compare the quality
of data. Unfortunately and unexpectedly  I notice that 
they disagree in the continuum limit even for the quenched case, as 
we will see.

In the main part the effect of dynamical quarks on hadron spectra
is discussed in the case of the Wilson quark action.
After the chiral extrapolation in the sea quark mass,
the effect becomes most visible, in particular on the hyper-fine
splitting of mesons, which consequently removes a large part of the 
quenching error in strange meson spectra.

In the last part, the string breaking, 
the topological susceptibility and the flavor-singlet meson mass
are discussed in connection with the dynamical quark effect.
This year mostly positive results are reported on all these
quantities.

Some important subjects such as quark masses, weak-matrix elements and
heavy quark physics in unquenched QCD will not be covered. Please 
see other reviews on these topics\cite{QM,WM,HQ}.

\begin{table}[tbh]
\caption{Summary of recent unquenched QCD simulations}
\label{tab:full}
\begin{tabular}{llll}
\hline
action & $a$(fm) & $La$ (fm) 
& $\mPS/\mV$ \\
\hline
$N_f = 2 $ & & & \\
PW\cite{SESAM}   &  0.086 &  1.4 & 0.69--0.83 \\
PW\cite{TCL}     & 0.086 & 2.0 & 0.55, 0.7 \\
PKS\cite{Columbia_full} & 0.09 & 1.5 & 0.57--0.70 \\
PKS\cite{MILC_full}   & 0.10--0.32 & 2.4--3.8 & 0.3--0.8 \\
PC(1.76)\cite{UKQCD_full}  & 0.12  & 1.0--1.9 & 0.67--0.86 \\
RC(TP)\cite{CPPACS_full}& 0.11--0.22 & 2.5--2.6 & 0.55--0.81 \\  
PC(NP)\cite{UKQCD_full_NP}   &  0.10 & 1.6 & 0.58--0.83 \\
PC(NP)\cite{JLQCD_full}   &  0.10 & 1.2--2.0 & 0.60--0.80 \\
\hline 
$N_f=3$ & & &  \\
SKS(I)\cite{MILC_3}          & 0.14 & 2.8   & 0.50--0.94 \\
\hline
\end{tabular}
\vspace{-0.8cm}
\end{table}
\section{Recent unquenched QCD simulations}
Recent large-scale unquenched 
simulations are summarized in Table~\ref{tab:full},
where $N_f$ is the number of flavors of dynamical quarks.
Throughout this paper, the following abbreviations are used to
distinguish different lattice actions in simulations.
The first letter stands for the gauge action:
P = Plaquette, R = RG improved\cite{iwasaki}, S = Symanzik improved\cite{LW},
while the second or more letters represent the quark action:
W = Wilson, KS = Kogut-Susskind, KS(I) = Improved KS\cite{naik,Asqtad}
C(X) = Clover\cite{SW} with \csw = X, where TP = Tadpole estimated value
and NP = Non-Perturbative value.
For example, PW stands for the plaquette gauge action and 
the Wilson quark action while
RC(TP) for the RG improved gauge action and the clover quark action with
the tadpole estimated value for \csw .

Simulations have been carried out at several lattice spacings
for the PKS\cite{MILC_full} and the RC(TP)\cite{CPPACS_full} actions.
Using data from these simulations one can make extrapolations to
the continuum limit.

\section{Several remarks on unquenched simulations}
In this section
several remarks are given on problems
associated with unquenched simulations.

\subsection{Auto-correlations}
\begin{figure}[tbh]
\centerline{\epsfxsize=7.5cm \epsfbox{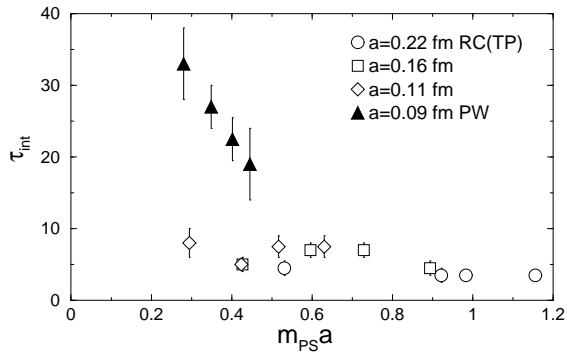}}
\vspace{-1cm}
\caption{Integrated auto-correlation time for the number of iterations,
$\tau_{int}^{N_{inv}}$ in the HMC algorithm, as a function of PS meson mass,
for the PW action\protect\cite{SESAM,TCL,SESAM_Auto} and 
the RC(TP) action\protect\cite{CPPACS_full}.}
\label{fig:auto}
\vspace{-0.5cm}
\end{figure}
Since unquenched simulations are very time-consuming, it is better to
perform hadron measurements as often as possible.
In such cases, one has to multiply statistical errors by $\sqrt{2\tau_{int}}$,
where $\tau_{int}$ is the integrated auto-correlation time of the
simulation, or equivalently, one should increase the bin-size of 
the Jack-knife analysis until the error reaches a plateau.

The integrated auto-correlation time for an observable ${\cal O}$
is defined as
\begin{equation}
\tau_{int}^{\cal O} =
\lim_{t\rightarrow\infty}\left[ \dfrac{1}{2}+\sum_{t'=1}^t 
\dfrac{C^{\cal O}(t')}{C^{\cal O}(0)} \right]
\end{equation}
where $C^{\cal O}(t)$ is the auto-correlation function for ${\cal O}$
given by
\begin{equation}
C^{\cal O}(t) = \langle {\cal O}(s) {\cal O}(s+t)\rangle_s
-\langle {\cal O}(s)\rangle_s \cdot\langle {\cal O}(s+t)\rangle_s 
\end{equation}
with $\langle F(s)\rangle_s \equiv \dfrac{1}{n}\sum_{s=1}^n F(s)$.

The auto-correlation time depends on the observable and, of course, the
simulation algorithm. In unquenched QCD,
$\pi$ meson propagator at a given time, or the number of iterations
for inversions of $D^\dagger D$ where $D$ is a lattice Dirac operator, 
seems to have a long auto-correlation time.
In Fig.~\ref{fig:auto}, $\tau_{int}$ for
the number of iterations $N_{inv}$ in the HMC algorithm is
plotted as a function of the PS meson mass, which corresponds to $\mPS/\mV
\ge 0.55$, in the case of Wilson-type (Wilson/Clover) quark actions
\cite{SESAM,TCL,CPPACS_full,SESAM_Auto}. 
We estimate $\tau_{int}^{N_{inv}} \simeq 5 \sim 40$, and
similarly  obtain $\tau_{int}^{\pi} \simeq 5 \sim 40$.
One should keep in mind that an accurate estimate of error for 
auto-correlation time requires $O(100-1000)\times \tau_{int}$ data,
so errors in these figures may well be under-estimated.
These $\tau_{int}$ are less or comparable to 
$\tau_{int}^\pi \simeq 40$ for the KS quark action\cite{MILC_full}.
It is noted that $\tau_{int}^{\pi,N_{inv}}$ for the PW are larger than
those for the RC action. This may be caused by
the difference of gauge actions, lattice spacings or
simulation parameters in the HMC algorithm.

From a practical point of view, $\tau_{int}\simeq 10$ is short enough, even
$\tau_{int}\simeq 40$ is not so bad, since recent unquenched simulations
can accumulate 8000 trajectories or more, which correspond to roughly
100--400 independent configurations.

It is expected that $\tau_{int}^{N_{inv}}$ increases as
the quark mass (PS meson mass) decreases, according to the scaling 
formula 
\begin{equation}
\tau_{int}^{N_{inv}}= A\cdot (\mPS a)^{-Z} .
\end{equation}
From data at $a \simeq 0.09$ fm for the PW action in the figure,
one obtains $Z =1.20(5)$\cite{SESAM_Auto},
which predicts a moderate number,
$\tau_{int}^{N_{inv}} \simeq 200 $, for $\mPS = 140$ MeV. 
Since this number is expected to be smaller for the RC action,
the auto-correlation may not be a major obstruction toward lighter
quark masses in unquenched simulations. 

It has been observed that 
the topological charge $Q$ has a long auto-correlation time
of the order of a few hundred trajectories or more
in unquenched HMC simulations  with the KS quark 
action.
This makes a measurement of the topological susceptibility
very difficult.
On the other hand, recent unquenched simulations 
give $\tau_{int}^Q \simeq 30-40$ for the RC(TP) action,
at $a=0.11$ fm and $\mPS/\mV =0.55-0.81$\cite{CPPACS_full}.
This value of $\tau_{int}^Q$ is small enough to make
the calculation of the topological susceptibility possible.
This difference may be explained by the fact that
the KS fermion determinant suppresses
the change of the topological charge at small quark masses
due to chiral symmetry.
Further investigations, however,
will be needed to confirm this speculation, for example, by employing the 
domain-wall quark action in unquenched simulations.

In conclusion the auto-correlation is not a major problem
for present unquenched simulations, except for the case of
the topological charge with the KS quark action\cite{IKMS}.
\label{sec:fse}
\begin{figure}[tbh]
\centerline{\epsfxsize=4cm \epsfbox{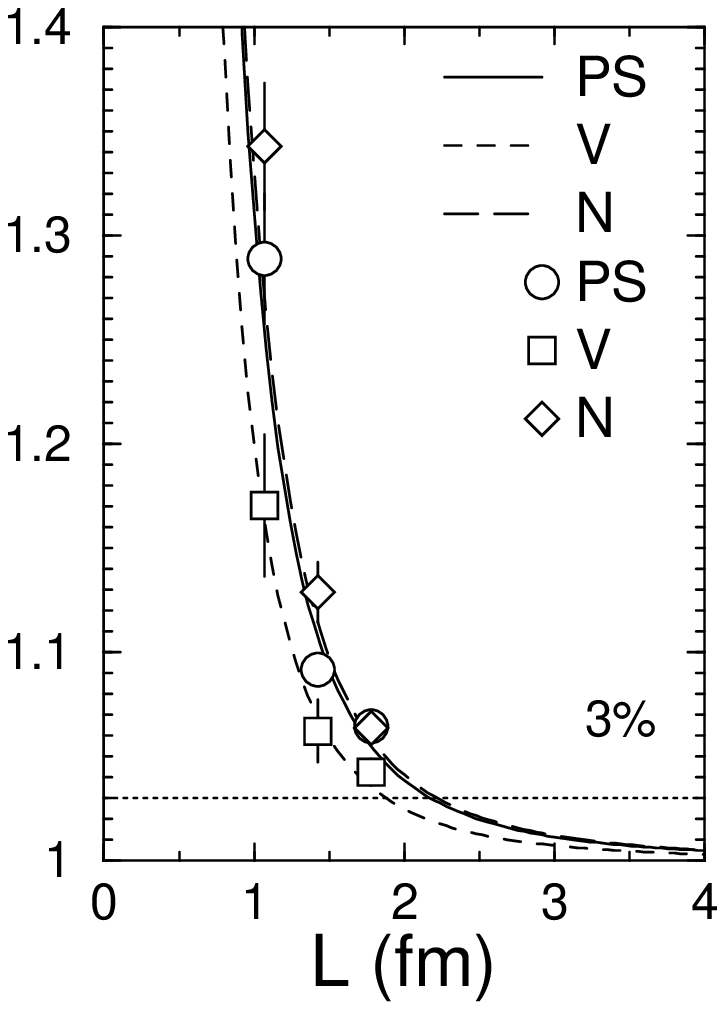}
            \epsfxsize=4cm \epsfbox{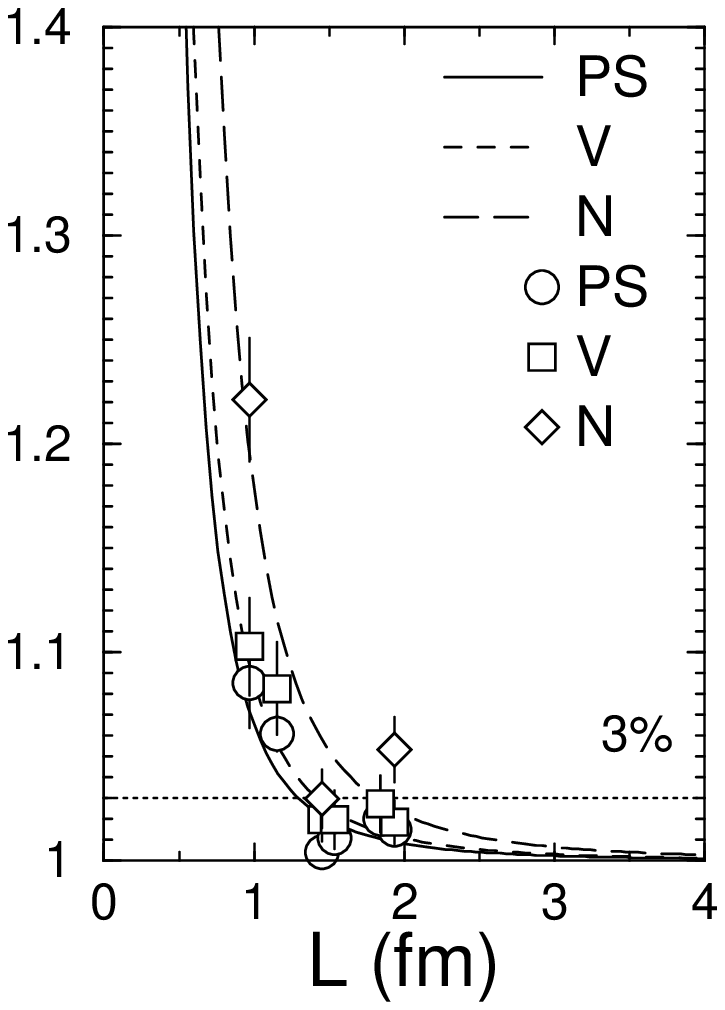} }
\vspace{-1cm}
\caption{Finite size errors on hadron masses $(\mH (L)-\mH)/\mH$ for
$H=$PS,V,N as a function of spatial extension $L$ (fm), in the case of
KS (left)\protect\cite{KEK_fse,MILC_fse}
and clover(right)\protect\cite{UKQCD_full,JLQCD_full} quark actions.}
\label{fig:fse}
\vspace{-0.5cm}
\end{figure}

\subsection{Finite size effects}
While finite size effects on hadron masses are known to be much
stronger in unquenched QCD than in quenched QCD for the KS quark 
action\cite{KEK_fse,MILC_fse},
recent unquenched simulations\cite{UKQCD_full,JLQCD_full}
suggest that finite size effects for
Wilson-type quark actions are milder.

To compare results from the two types of actions directly, 
hadron masses at a finite extension $L$ are fitted by the form
\begin{equation}
\mH(L) = \mH + \dfrac{c}{L^3},
\end{equation}
which is found to be good for the range of $L$ in the current 
simulations\cite{KEK_fse}, and therefore is used as a working hypothesis 
for the comparison.
Relative errors, $(\mH(L)-\mH)/\mH$, for H= PS(Pseudo Scalar), V(Vector) and 
N(Nucleon), are plotted as a function of $L$ in Fig.~\ref{fig:fse}, 
at $a\simeq 0.09$ fm and $\mPS/\mV \simeq 0.68$
for the KS quark action\cite{KEK_fse,MILC_fse} 
and at $a\simeq 0.12$ fm and $\mPS/\mV \simeq 0.72$ for the clover quark 
action\cite{UKQCD_full,JLQCD_full}.
At similar lattice spacings and $\mPS/\mV$,
the size effect seems to be larger for the KS quark. 

\begin{table}[htb]
\caption{Lattice size $L$(fm) which give 3\% relative errors.}
\label{tab:fse}
\begin{tabular}{lllll}
\hline
$\mPS/\mV$ & $a$(fm) & PS & V & N \\
\hline
\multicolumn{2}{c}{KS}& & & \\
0.5 & 0.2 & 1.2 & 1.9 & 2.2 \\
    & 0.09 & 2.6 & 2.2 & 2.6 \\
0.68 & 0.09 & 2.2 & 2.0 & 2.3 \\
\multicolumn{2}{c}{Wilson-type}& & & \\
0.72 & 0.12 & 1.4 & 1.5 & 1.8 \\
0.75 & 0.12 & 1.0 & 1.4 & \\
0.8 & 0.12  & 1.1 & 1.3 & 1.7\\
\hline
\end{tabular}
\vspace{-0.5cm}
\end{table}

Using data currently available, 
the lattice size $L$ (fm) which gives a 3\% relative finite size error, 
$(\mH(L)-\mH)/\mH \simeq 0.03$, is summarized in Table~\ref{tab:fse}.

From Fig.\ref{fig:fse} and table \ref{tab:fse}, 
one observes that the size effect is larger for
the KS quark action than for the Wilson-type action.
Moreover patterns of the size effect are different:
size effects of PS and N are comparable and larger than that of V for the KS
quark action while the size effect of N dominates for the Wilson-type quark 
action. 
This difference is a part of scaling violation since 
finite size effect is universal in the continuum limit,
and it may be caused by the chiral properties of
the two actions: 
the KS quark retains chiral symmetry while the Wilson-type quark 
breaks it explicitly.
Another possibility is that the ``effective'' lattice size for the KS quark
may be smaller than the lattice size for the Wilson quark, due to the
spread of the KS quark over a hyper-cube in the spinor-flavor interpretation.
Obvious disadvantages of Wilson-type quarks such as the absence of chiral 
symmetry, $O(a)$ scaling violation and larger costs of numerical 
simulations are partly compensated by this milder finite size effect.

\subsection{Common scale}
\begin{figure}[tbh]
\centerline{\epsfxsize=3.6cm \epsfbox{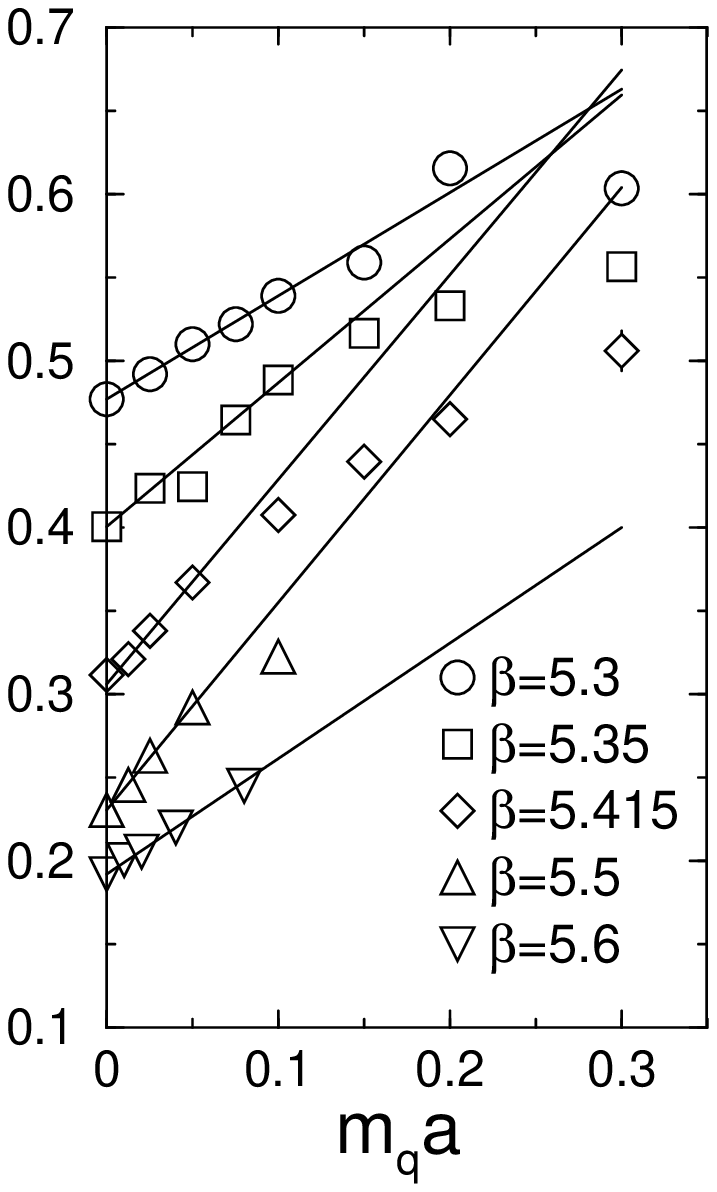}
            \epsfxsize=3.8cm \epsfbox{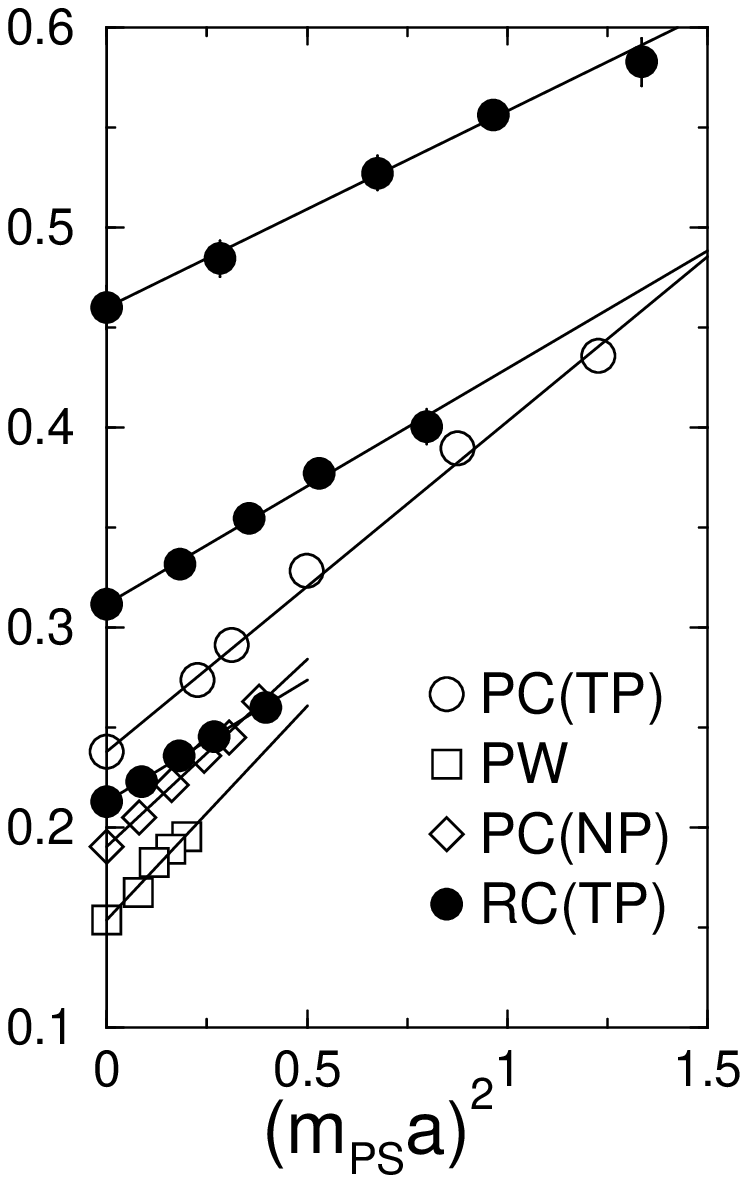} }
\vspace{-1cm}
\caption{(Left) The quark mass dependence of the inverse of the Sommer scale
for the PKS action\protect\cite{MILC_r0}.
(Right) The same for PC(TP)\protect\cite{UKQCD_full}, 
PW\protect\cite{SESAM-TCL_r0}, RC(NP)\protect\cite{JLQCD_full} and
RC(TP)\protect\cite{CPPACS_full} actions.
Lines are fits as explained in the text.}
\label{fig:r0}
\vspace{-0.5cm}
\end{figure}
One has to fix a scale to compare results in different simulations.
In quenched simulations, it may be easier and better to use a scale 
determined only by the gauge action such as the string tension.
In unquenched simulations, however,
all quantities depend on quark actions and quark masses as well as
the gauge action and the gauge coupling constant.

For the common scale in unquenched simulations,
one had better to use the Sommer scale $r_0$\cite{sommer}, which is given by 
\begin{equation}
r_0^2 \dfrac{d\ V(r)}{d\ r}\vert_{r=r_0} = 1.65 ,
\end{equation}
where $V(r)$ is the static quark potential.

This quantity, estimated as $r_0\simeq 0.49$ fm,
is insensitive to either the short-distance physics of QCD 
such as the Coulomb term or the long distance one represented by string 
breaking.
In simulations $r_0$ is accurately determined by numerical derivative of
the potential\cite{sommer} or by the relation $r_0^2=(c+1.65)/\sigma$,
where the string tension $\sigma$ and the Coulomb coefficient $c$ are
extracted from the fit of $V(r)$.
Two methods give consistent values of $r_0$
with reasonably small errors\cite{MILC_3}.

The quark mass dependence of the Sommer scale is
shown in Fig.\ref{fig:r0},
where the vertical axis is the inverse of the Sommer scale, while
the horizontal one is the bare quark mass for the KS quark action\cite{MILC_r0}
or the PS meson mass squared for Wilson-type quark 
actions\cite{UKQCD_full,CPPACS_full,JLQCD_full,SESAM-TCL_r0}.
One can easily obtain the value of $r_0$ in the chiral limit,
fitting the data for small enough quark masses by the linear form
\begin{equation}
\dfrac{a}{R_0} =\dfrac{a}{r_0} + A\cdot M ,
\end{equation}
where $M=  m_q a$ or $\mPS^2a^2$.

Later the Sommer scale at $M\not=0$ becomes necessary.
For distinction I use $R_0$ in such cases instead of $r_0$, which stands
for exclusively the value in the chiral limit.

\subsection{Changing quark mass}
In unquenched simulations, it is non-trivial to change
the quark mass while keeping the lattice spacing constant. 
In a standard method one varies the bare quark mass $m_q a$ at fixed 
$\beta$\cite{SESAM,TCL,Columbia_full,MILC_full,UKQCD_full,CPPACS_full,JLQCD_full}, 
determining the lattice spacing in the chiral limit or 
at the physical point of $\mPS/\mV = 0.135/0.77$.
In recent simulations\cite{UKQCD_full_NP,MILC_3}
one employs a new method in which both $m_q a$ and $\beta$ are varied to
keep $R_0$ constant, claiming that the lattice spacing 
becomes smaller at smaller quark masses otherwise.
In this subsection I will discuss if this claim is true.

First of all it is noted that the lattice spacing at different quark masses
can not be compared in principle since no experimental input is available away 
from the physical quark mass.

Let me discuss a relation between $a$ and
bare parameters in QCD using the running coupling constant $\bar g^2 (L)$ 
instead of $R_0$ to see a difference of two methods,
since the perturbative expansion in an $O(a)$ improved theory 
is available for $\bar g^2 (L)$ as\cite{ALPHA}
\begin{eqnarray}
\bar g^2(L) &=& g_0^2 + g_0^4 \left[\beta_0 \log (L/a) + f (m_R L) \right.
\nn\\
& & + \left. k\ m_R a\right] + O(a^2, g_0^6) ,
\end{eqnarray}
where $g_0^2$ is the bare coupling constant, $L$ is the lattice size, 
$m_R$ is the renormalized quark mass.

The mass-independent renormalization scheme is defined through the
renormalized coupling constant $g_R^2$, defined by
\begin{equation}
g_R^2 = g_0^2[1-g_0^2\beta_0\log (\mu a)]+ O(g_0^6),
\end{equation}
where $\mu$ is the renormalization scale. 
This scheme can be easily related to the $\overline{MS}$ scheme.
Clearly the standard method of unquenched simulations corresponds to 
this scheme, since $a$ depends only on $g_0^2$ for a fixed $g_R^2$.
The running coupling becomes
\begin{eqnarray}
\bar g^2(L) &=& g_R^2 + g_R^4\left[\beta_0 \log (L\mu) + f (m_R L) \right.\nn\\
& & + \left. k\ m_R a\right] + O(a^2, g_R^6),
\end{eqnarray}
which has $m_R a$ scaling violation even in the $O(a)$ improved theory,
in addition to the physical mass dependence $f(m_R L)$.
Therefore the analogy suggests that $R_0$ has scaling violation of $m_R a$ 
as well as the physical quark mass dependence in this scheme.

The mass-dependent scheme is specified by
\begin{eqnarray}
\hat g_{R}^2 &=& g_0^2[1+g_0^2\{-\beta_0\log (\mu a)+f(m_R L)  \nn \\
& & + k\ m_q a\}] + O(g_0^6),
\end{eqnarray}
where $m_q$ is the bare quark mass. The lattice spacing
$a$ depends on both $g_0^2$ and $m_q a$ for a fixed $\hat g_R^2$,
in such a way that the running coupling
becomes  mass-independent:
\begin{equation}
\bar g^2(L) = \hat g_{R}^2 + \hat g_{R}^4\beta_0 \log (L\mu)+O(a^2, 
\hat g_{R}^6).
\end{equation}
Up to $O(a^2)$  ambiguity one can define the relation
between $g_0^2$ and $m_q a$ so that $\bar g^2(L)$
instead of $\hat g_R^2$ becomes constant.
In other words the new method of unquenched simulations
corresponds to this renormalization scheme,
which  keeps $O(a)$ improvement in $\bar g^2$
while has a complicated relation to $\overline{MS}$ scheme.

This consideration concludes that two methods of changing quark mass in
unquenched simulations correspond to different renormalization schemes
and that the lattice spacing is kept constant in both methods, contrary to the
claim of ref.\cite{UKQCD_full_NP,MILC_3}.

Ref.~\cite{ALPHA} has proposed another renormalization scheme, called an 
improved mass-independent scheme, whose renormalized coupling constant is 
defined by
\begin{equation}
\tilde g_{R}^2 = 
\tilde g_0^2[1-\tilde g_0^2\beta_0\log (\mu a)]+ O(\tilde g_0^6)
\end{equation}
where $\tilde g_0^2 =( 1 + b_g m_q a) g_0^2$ is called an improved bare
coupling constant with $b_g = k g_0^2 + O(g_0^4)$.
In this scheme 
$a$ depends on both $g_0^2$ and $ m_q a$ through $\tilde g_0^2$
for a fixed $\tilde g_R^2$,
while the relation of this scheme to the $\overline{MS}$ scheme remains
simple since the mass-dependence in $\tilde g_0^2$ vanishes as  $O(a)$.
The running coupling constant in this scheme has a mass-dependence and
is $O(a)$ improved:
\begin{equation}
\bar g^2(L) = \tilde g_R^2 + \tilde g_R^4[\beta_0 \log (L\mu) + f (m_R L)] +
 \cdots .
\end{equation}

Let me close this section by suggesting two options for unquenched
simulations in the future.
One can keep using the standard mass-independent scheme with the special care
about $m_q a$ scaling violation, which may not be so large for
light and strange quarks.
One can instead try the improved mass-independent scheme with either
the perturbative estimate for $b_g$, where the 1-loop value is
available\cite{SS} or the non-perturbative estimate for $b_g$,
which seems possible by the method of ref.~\cite{MRSSTT}.

\section{Hadron spectra: Wilson vs. KS quarks}
In order to disentangle the dynamical quark effect from the scaling violation,
one has to compare unquenched results with quenched ones
in the continuum limit.
Although precise quenched hadron spectra
have already been obtained with the PW
action\cite{CPPACS_quench}, 
it is better to confirm the results with different actions.
Such a comparison becomes more important in unquenched QCD,
where it is much harder to control the continuum extrapolation.
Therefore, in this section, hadron spectra in the continuum limit are 
systematically compared 
between Wilson and KS quark actions.

\subsection{Edinburgh plot in quenched QCD}
\begin{figure}[tbh]
\centerline{\epsfxsize=7.5cm \epsfbox{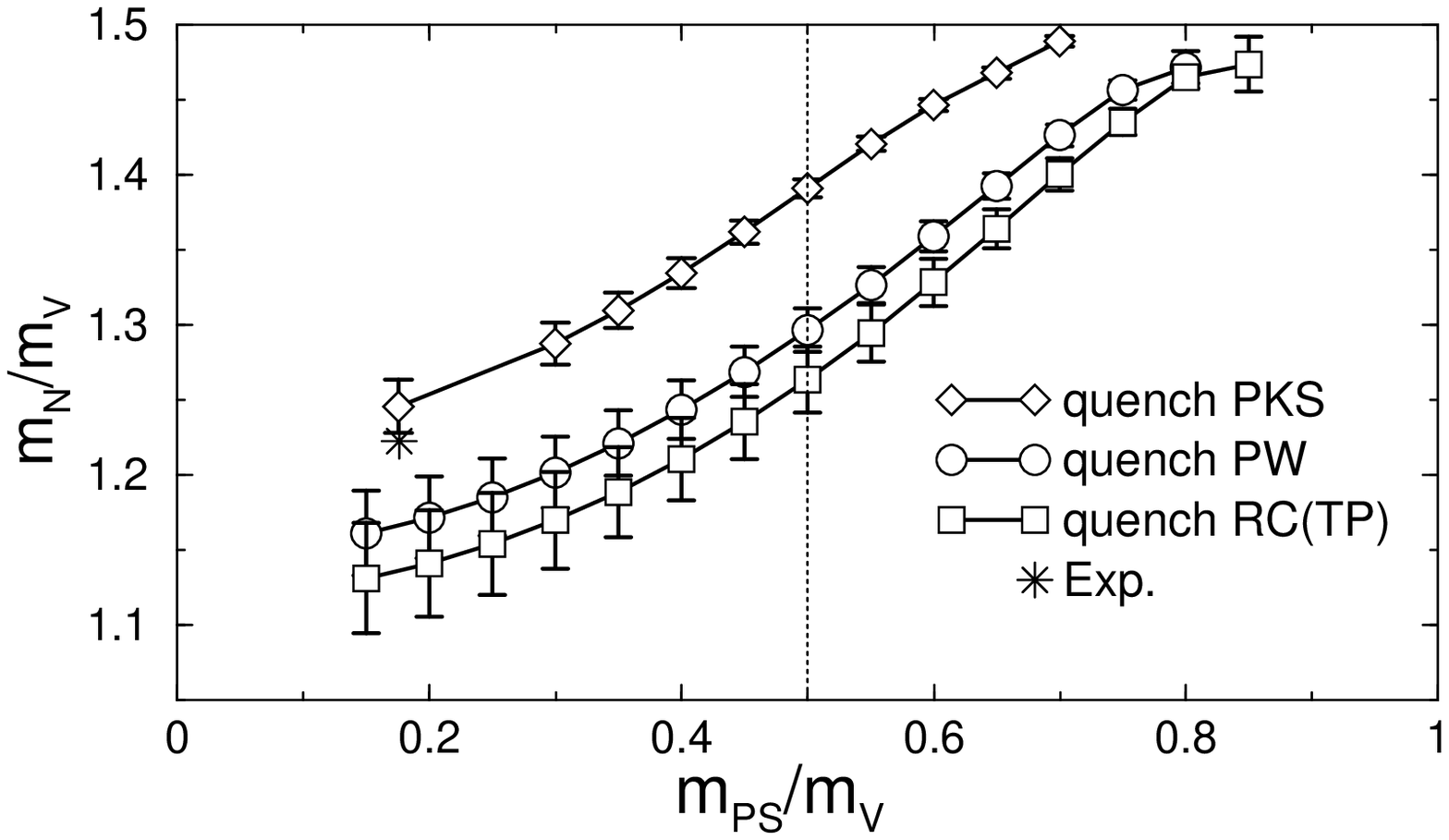}}
\centerline{\epsfxsize=7.5cm \epsfbox{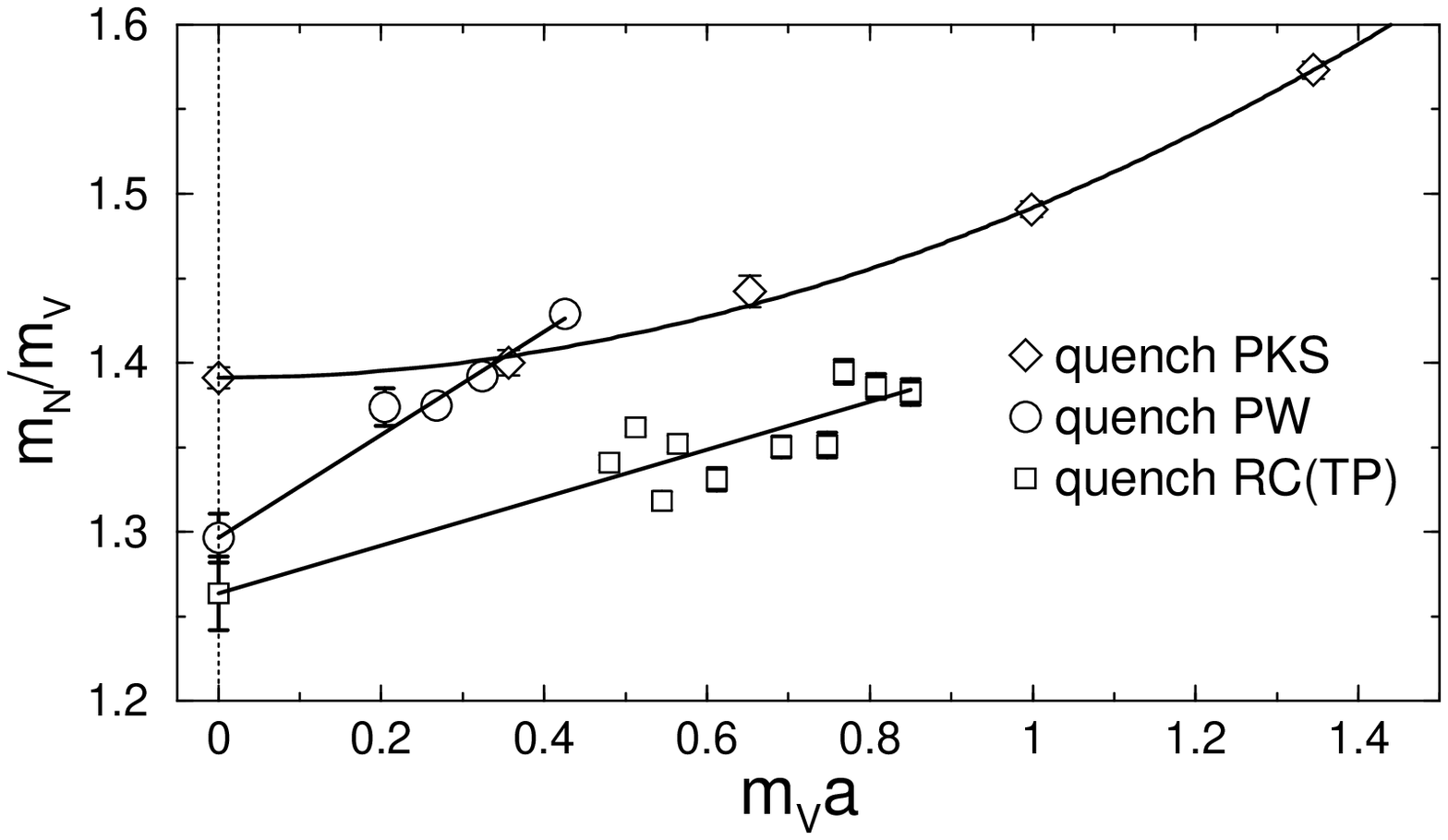}}
\vspace{-1cm}
\caption{(Upper) The Edinburgh plot($\mN/\mV$ vs $\mPS/\mV$)
in quenched QCD at $a=0$ for 
PKS\protect\cite{MILC_quench}, PW\protect\cite{CPPACS_quench}
and RC(TP)\protect\cite{CPPACS_full} actions.
(Lower) $\mN/\mV$ as a function of $\mV a$ at $\mPS/\mV = 0.5$
for same actions.}
\label{fig:edinN_quench}
\vspace{-0.5cm}
\end{figure}
In the continuum limit of quenched QCD 
the nucleon mass is found to be a little smaller
than the experimental value for the PW 
action\cite{CPPACS_quench} while it is consistent with the
experimental value for the KS action\cite{MILC_quench}.
To see if this small difference is caused by the 
ambiguity associated with the chiral extrapolation,
the Edinburgh plot($\mN/\mV$ vs. $\mPS/\mV$) in the continuum limit
of quenched QCD is shown in the upper part of Fig.~\ref{fig:edinN_quench},
for PKS\cite{MILC_quench}, PW\cite{CPPACS_quench} and 
RC(TP)\cite{CPPACS_full} actions.

Surprisingly results between Wilson and KS quark actions 
in the continuum limit of quenched QCD
do not agree at all in the whole range of quark masses.
It should be noted, however, that 
the Edinburgh plot from the PW action
is consistent with the one from the RC(TP) action,
ensuring an agreement among Wilson-type actions in the 
continuum limit.

Neither the finite size effect nor the chiral extrapolation can be 
a main source for this discrepancy, since the volume is large enough
($L a\simeq$ 3 fm for PW, $2.6\sim 3$ fm for PKS) and the
discrepancy is significant in the whole range of quark masses.

To investigate the quality of the continuum extrapolation,
$\mN/\mV$ is shown in the lower part of Fig.~\ref{fig:edinN_quench}
as a function of the lattice spacing $\mV a$ at 
$\mPS/\mV = 0.5$, denoted by the vertical line in the Edinburgh plot,
where both finite size effect and ambiguity of the chiral extrapolation
should be small.

Although the quenched PW result has larger scaling violation than the
quenched RC(TP) result, they agree in the continuum limit,
concluding that the continuum extrapolation is controlled
for Wilson-type quark actions.
The scaling violation of the quenched PKS result is well fitted quadratically
in $\mV a$, which is a leading behavior theoretically expected
for the KS quark action\cite{sharpe}.
It is therefore clear that the discrepancy between quenched results
is not caused by the  continuum extrapolations.
I have to conclude that the discrepancy remains unexplained.

\subsection{Edinburgh plot in unquenched QCD}
\begin{figure}[tbh]
\centerline{\epsfxsize=7.4cm \epsfbox{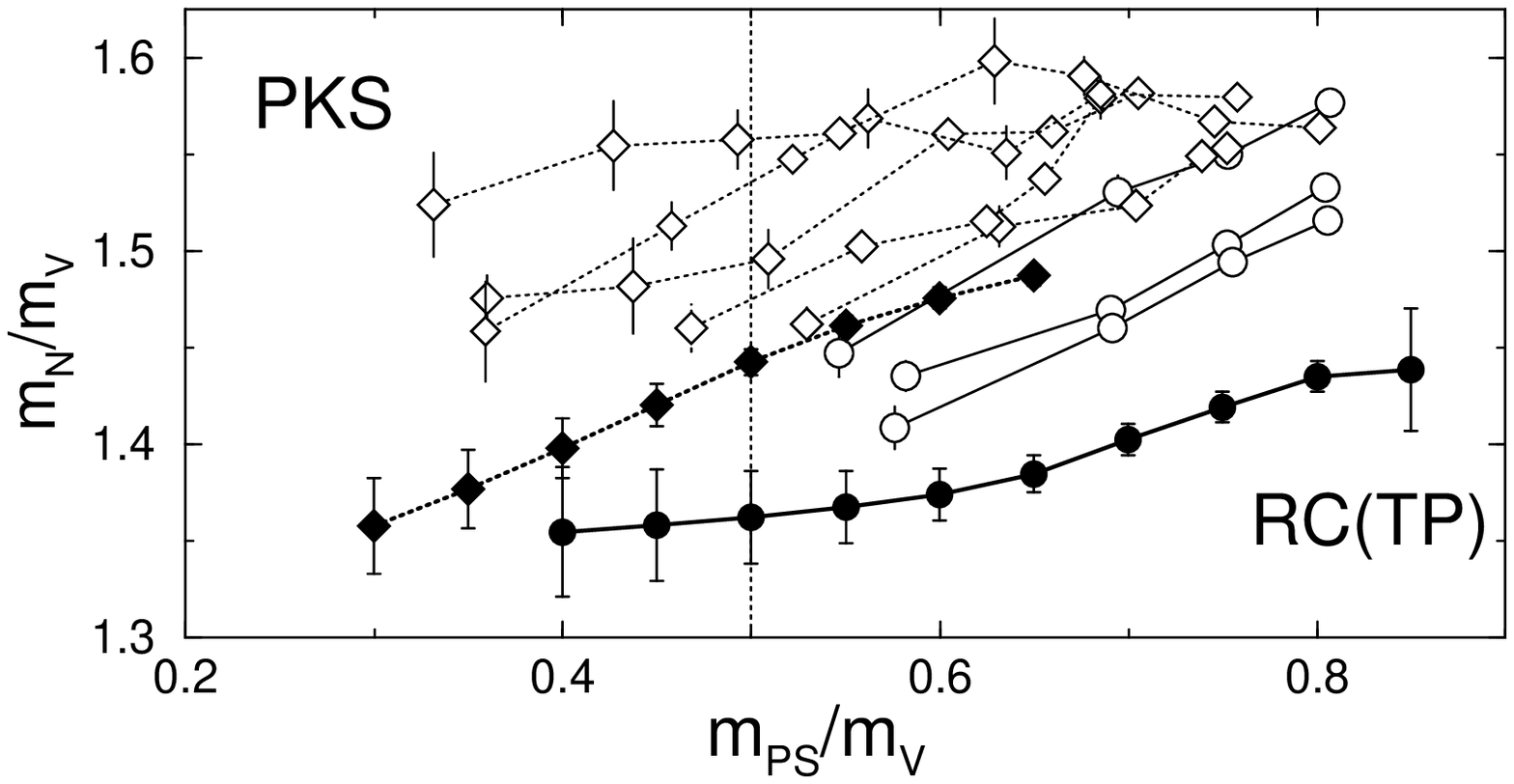}}
\centerline{\epsfxsize=7.4cm \epsfbox{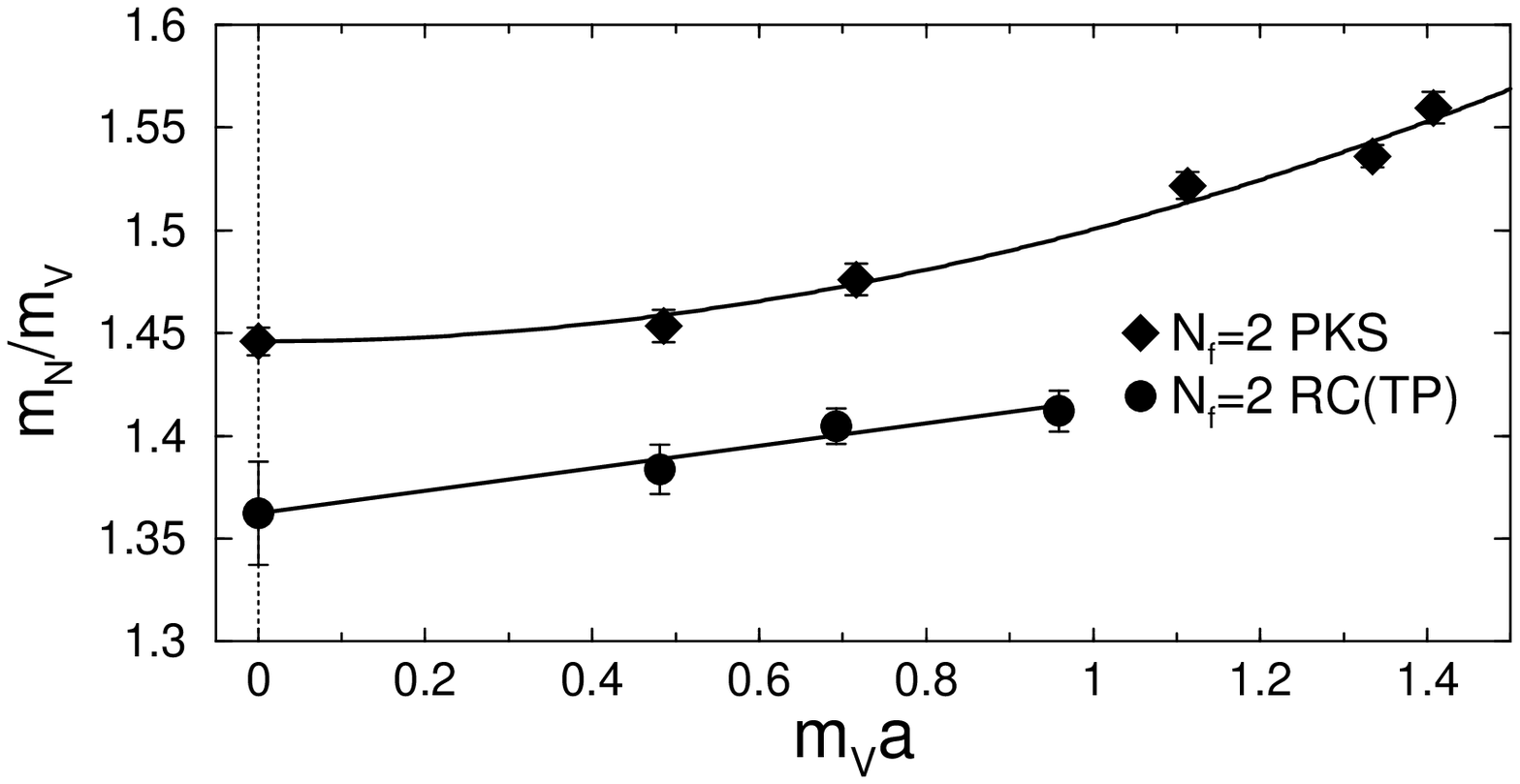}}
\vspace{-1cm}
\caption{(Upper) The Edinburgh plot in unquenched QCD for 
the PKS action(diamonds)\protect\cite{MILC_full} and 
the RC(TP) action(circles)\protect\cite{CPPACS_full}.
Open symbols represent data at 
$a =$ 0.32, 0.28, 0.23, 0.15, 0.10 fm for the former and
0.22, 0.16, 0.11 fm for the latter, from above to below,
while solid symbols correspond to results in the continuum limit.
(Lower) $\mN/\mV$ as a function of $\mV a$ at $\mPS/\mV = 0.5$
for same actions.}
\label{fig:edinN_full}
\vspace{-0.5cm}
\end{figure}

The analysis in the previous subsection is applied also to unquenched results
for PKS\cite{MILC_full} and RC(TP)\cite{CPPACS_full} actions,
where the finite size effect is expected to be significant,
in particular for lighter quark masses.
The upper part of Fig.~\ref{fig:edinN_full} shows
the Edinburgh plot in the continuum limit,
which gradually flattens for the RC(TP)\cite{CPPACS_full} 
while continuously decreases for the PKS\cite{MILC_full}
toward lighter quark masses, together with results directly obtained
by simulations at non-zero $a$ without extrapolations.
The behavior for the RC(TP) indicates the existence of the finite size
effect as well as the difficulty of the chiral extrapolation from
the restricted range of quark masses such that $\mPS/\mV \ge 0.55$.
On the other hand, the behavior for the PKS, which is irregular 
at non-zero, in particular larger $a$, makes 
both chiral and continuum extrapolations difficult.
Consequently, the quality of 
data at lighter quark masses for both actions is
insufficient for the reliable continuum  extrapolation in
the Edinburgh plot.
Therefore I concentrate on results at heavy masses,
where the disagreement between the two actions is already manifest.

The lower part of Fig.~\ref{fig:edinN_full}
shows $\mN/\mV$ as a function of the lattice spacing $\mV a$ at 
$\mPS/\mV = 0.5$, which
is obtained by the interpolation for
the PKS action or by a little extrapolation for the RC(TP) 
action.
The scaling violation is not so significant for both
RC(TP) and PKS actions. In particular the result for the latter
is well fitted quadratically in $\mV a$.
Two results largely deviate even at non-zero $a$, and the difference
stays almost constant toward the continuum limit.
Therefore the deviation is not brought into by the continuum extrapolation,
though the linear continuum extrapolation for PC(TP) data
might be affected a little by higher order terms such as $a\log a$ or $a^2$.
The lattice size, $L a\simeq 2.5$ fm for the RC(TP) or 2.4 fm for the PKS,
may not be large enough at $\mPS/\mV=0.5$ in unquenched QCD.
As mentioned in Sect.\ref{sec:fse}, there may be 3\% error in $\mN$ for
the KS quark action at $L a \simeq$ 2.6 fm and $a\simeq 0.1$ fm.
Although no data are available for the Wilson quark action at this
$\mPS/\mV$, the error is expected to be smaller than the KS quark action.
With all systematics together, it may be too early to conclude
that results between Wilson and KS quark actions
disagree in the continuum limit at $\mPS/\mV=0.5$. Further investigations,
in particular on the finite size effect or at larger mass ratios such that
$\mPS/\mV=0.6$ to avoid the chiral extrapolation, 
are urgently needed to draw definite conclusions.

\subsection{Hadron spectra normalized by $R_0$}
\begin{figure}[tbh]
\centerline{\epsfxsize=7.5cm \epsfbox{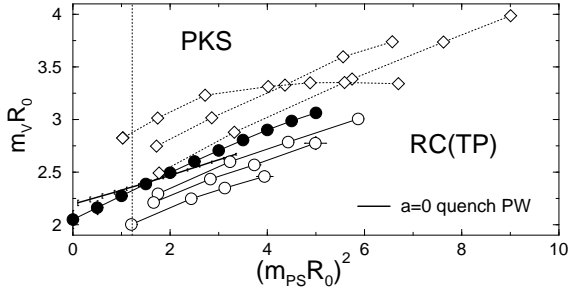}}
\vspace{-1cm}
\caption{$\mV R_0$ as a function of $(\mPS R_0)^2$ in
unquenched QCD for the PKS action(diamonds)\protect\cite{MILC_full}
and the RC(TP) action(circles)\protect\cite{CPPACS_full}.
Open symbols represent data at non-zero $a$, while
solid ones in the continuum limit.
Solid line represents quenched data in the continuum limit 
for the PW action\cite{CPPACS_quench} for comparison.}
\label{fig:vps2}
\vspace{-0.5cm}
\end{figure}
In the previous subsection, the discrepancy in the Edinburgh plot 
between Wilson and KS quark actions became manifest in the continuum limit, 
in particular, for the quenched case.
However, since the Edinburgh plot represents the relation among mass ratios,
it is difficult to find a source responsible for the discrepancy.
Alternatively, in this subsection,
hadron masses normalized by the Sommer scale
are considered as a function of $(\mPS R_0)^2$, which
plays the role of quark mass. 
With this normalization one can keep the simple
relation between $\mH$ and  $\mPS^2$ 
as long as the quark mass dependence of $R_0$ is sub-leading,
while still avoiding the chiral extrapolation.

In Fig.~\ref{fig:vps2} $\mV R_0$ from unquenched simulations is plotted 
as a function of $(\mPS R_0)^2$ at non-zero $a$, for both 
RC(TP)\cite{CPPACS_full} and PKS\cite{MILC_full} actions,
together with lines in the continuum limit from 
quenched PW\cite{CPPACS_quench} and unquenched RC(TP).
Unfortunately lines in the continuum limit  are not available for
the PKS action.

Data at non-zero $a$ approach the continuum limit
from below($a=$ 0.22, 0.16, 0.11 fm) for RC(TP) and from 
above($a=$ 0.21, 0.15, 0.11 fm) for PKS. 
The behavior looks reasonable except for PKS at $a\simeq 0.21$ fm.
Roughly speaking slopes are similar among unquenched results
and steeper than the quenched one.
\begin{figure}[tbh]
\centerline{\epsfxsize=3.7cm \epsfbox{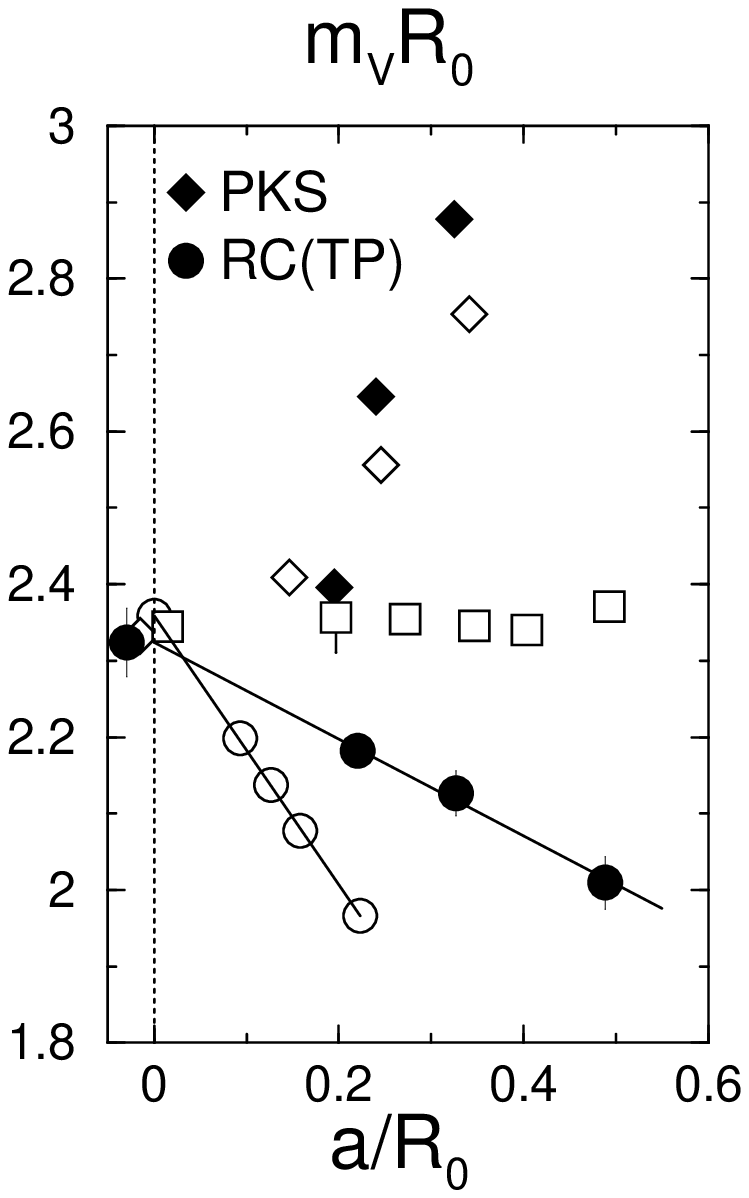}
            \epsfxsize=3.7cm \epsfbox{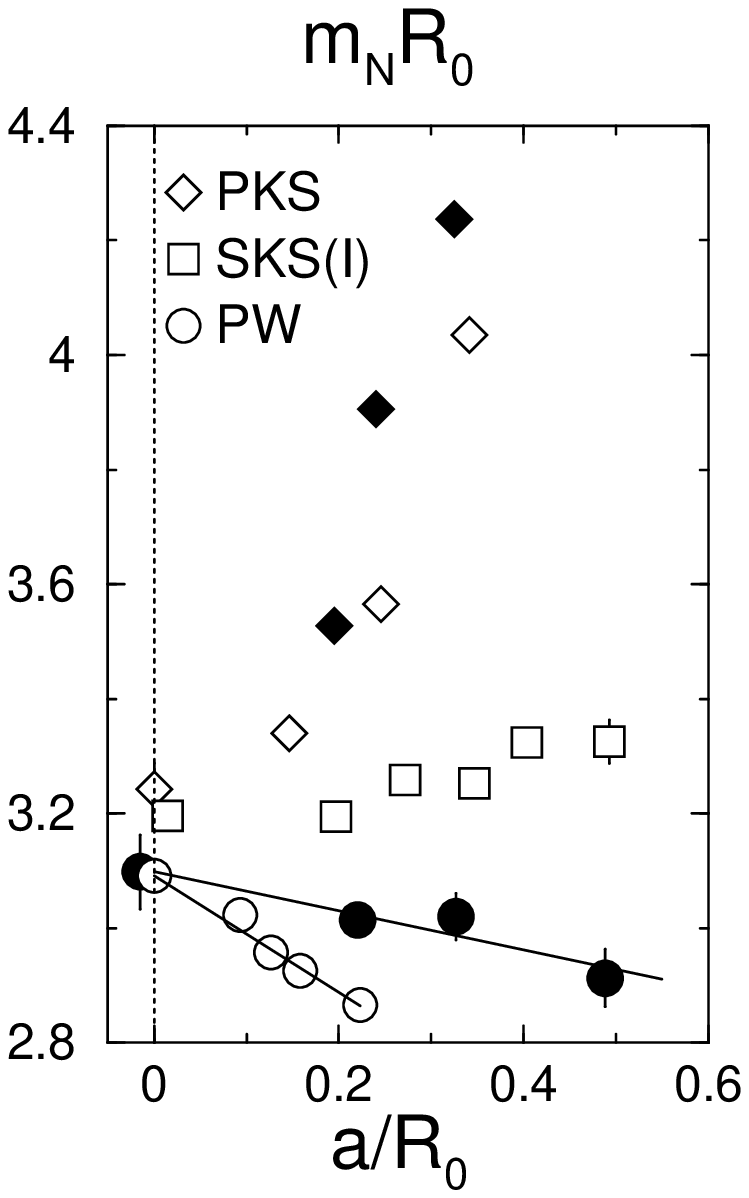}}
\vspace{-1cm}
\caption{$\mV R_0$(left) and $\mN R_0$(right) as a function of $a/R_0$
at $(\mPS R_0)^2$ = 1.22, in quenched QCD(open symbols) with 
PW\protect\cite{CPPACS_quench}, PKS\protect\cite{MILC_quench} and 
SKS(I)\protect\cite{MILC_imp,MILC_decay} actions and
in unquenched QCD(solid symbols) with RC(TP)\protect\cite{CPPACS_full} and 
PKS\protect\cite{MILC_full} actions.}
\label{fig:vnr0@1.22}
\vspace{-0.5cm}
\end{figure}

As before the scaling behavior is examined at $(\mPS R_0)^2 $=1.22,
where PKS data are available. 
In the left part of Fig.~\ref{fig:vnr0@1.22} $\mV R_0$ at $(\mPS R_0)^2 $=1.22
is shown as a function of $a/R_0$.
The quenched PKS data\cite{MILC_quench} and SKS(I) 
data\cite{MILC_imp,MILC_decay}
agree with each other in the quadratic continuum extrapolation, and 
an extremely good scaling behavior is observed for the latter.
The continuum values also agree with that for the PW action.
Therefore no discrepancy is found for quenched vector meson masses between
KS and Wilson quarks.

Similarly
the scaling behavior of $\mN R_0$ 
is shown in the right part 
of Fig.~\ref{fig:vnr0@1.22} at $(\mPS R_0)^2 $=1.22.
The linear continuum extrapolation of quenched PW data\cite{CPPACS_quench} 
gives a smaller value than the quadratic one of quenched PKS 
data\cite{MILC_quench},
which well agrees with the one of SKS(I) data\cite{MILC_imp,MILC_decay}.

The individual analysis for $\mV$ and $\mN$ suggests that
a discrepancy in $\mN$ 
is a main source for the disagreement in the Edinburgh plot between 
PW and PKS actions in quenched QCD.

For unquenched QCD,
continuum extrapolations of $\mV$ and $\mN$ are reasonably good 
for the RC(TP) action\cite{CPPACS_full} and 
agree with those for quenched QCD with PW\cite{CPPACS_quench},
while very large scaling violations for $\mV$ and $\mN$,
which do not show expected $a^2$ behaviors, make continuum 
extrapolations of individual masses impossible 
for the PKS action\cite{MILC_full}.
However these large scaling violations in $\mV$ and $\mN$ 
cancel each other in the ratio $\mN/\mV$ as seen
in the lower part of Fig.~\ref{fig:edinN_full}.

\subsection{Suggestions on further investigations}
From available data so far we can not identify a definite
reason for the discrepancy in the Edinburgh plot between Wilson and KS quark 
actions, in particular for quenched QCD, where
the disagreement is found only in $\mN$.
To completely resolve the problem, we should accumulate
as many data as available for all varieties of
actions and simulations.

We should analyze the Edinburgh plot and the plot of $\mH R_0$ vs 
$(\mPS R_0)^2$, as well as their scaling behavior
at $\mPS/\mV $=0.5, 0.7 for the former and at $(\mPS R_0)^2$ = 1.22, 3.0
for the latter as typical values  already used in several 
literatures.

Even quenched results from the third type of quark actions such as
domain-wall or overlap will become useful to resolve the problem.

\section{Dynamical quark effects on hadron spectra}
In this section, dynamical quark effects on hadron spectra are investigated,
comparing data from unquenched simulations with those from quenched ones. 
Due to the discrepancy between Wilson and KS quark actions 
in the previous section,
results only for Wilson-type quark actions
are considered here, since 
more data are available so far.

\subsection{Quark mass dependence in partially quenched analysis}
In unquenched simulations, valence quark masses can be different from
the mass of dynamical quarks (sea quark mass).
In such a partially quenched analysis,
the PS meson mass squared are employed for Wilson-type quark actions
to specify how large the quark mass is,
without chiral extrapolation necessary to define the quark mass
through the critical hopping parameter.

The following equality is then assumed for the PS meson mass.
\begin{eqnarray}
\mPSv^2 &\equiv&
\dfrac{1}{2}\left[\mPS^2(s;v_1,v_1)+\mPS^2(s;v_2,v_2)\right]\nn \\
&=& \mPS^2(s;v_1,v_2)
\label{eq:ps-ps}
\end{eqnarray}
where $s$ specifies a sea quark mass for degenerate $N_f=2$ dynamical quarks
and $v_1$ and $v_2$ denote valence quark masses in the meson.
This assumption is correct at the lowest order of chiral perturbation theory
and is well-satisfied numerically in data of ref.~\cite{JLQCD_full}.
\begin{figure}[tbh]
\centerline{\epsfxsize=7.5cm \epsfbox{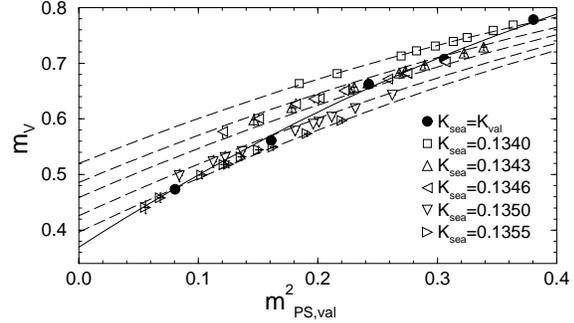}}
\vspace{-1cm}
\caption{$\mV(s;v_1,v_2)$ in lattice units
as a function of $\mPSv^2$ at $\beta = 5.2$ for the PC(NP) action 
on a $16^3\times 48$ lattice\protect\cite{JLQCD_full}.
}
\label{fig:vps2_jlqcd}
\vspace{-0.5cm}
\end{figure}

In Fig.\ref{fig:vps2_jlqcd}
$\mV(s;v_1,v_2)$ is plotted 
in lattice units as a function of $\mPSv^2$\cite{JLQCD_full}.
At a fixed sea quark mass($K_{sea}=0.1355$, for example, which
corresponds to $\mPS/\mV \simeq 0.6$), data lie on one curve with
some curvature, suggesting that higher order terms
in $\mPSv^2$ are necessary while the effect of
$(\mPS^2(s;v_1,v_1)-\mPS^2(s;v_2,v_2))^2$ term is negligible.
Therefore, as shown in the figure,
data can be reasonably fitted by the form
\begin{eqnarray}
\mV(s;v_1,v_2)&=& A + B_s\cdot \mPSs^2 + C_s\cdot (\mPSs^2 )^2\nn\\
&+&(B_v + C_{sv}\cdot \mPSs^2)\cdot \mPSv^2 \nn \\
&+& C_v\cdot (\mPSv^2)^2
\label{eq:vector}
\end{eqnarray}
with $\mPSs^2=\mPS^2(s;s,s)$, where all possible terms of $O(\mPS^4)$
are included.
The fact that the intercept and the curvature of each curve depend on the 
sea quark mass
might be interpreted as a dynamical quark effect. 

\begin{figure}[tbh]
\centerline{\epsfxsize=3.8cm \epsfbox{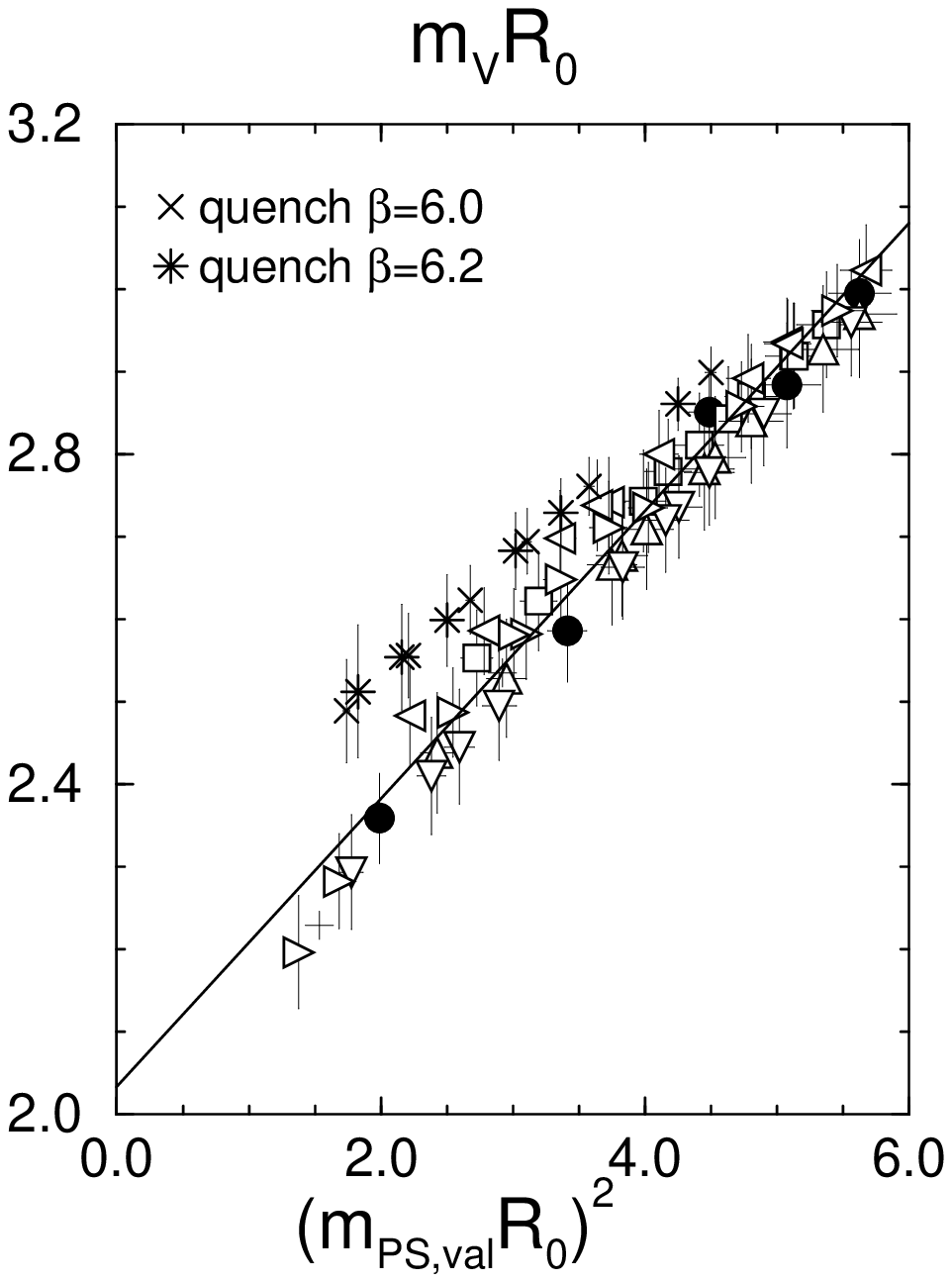}
            \epsfxsize=3.8cm \epsfbox{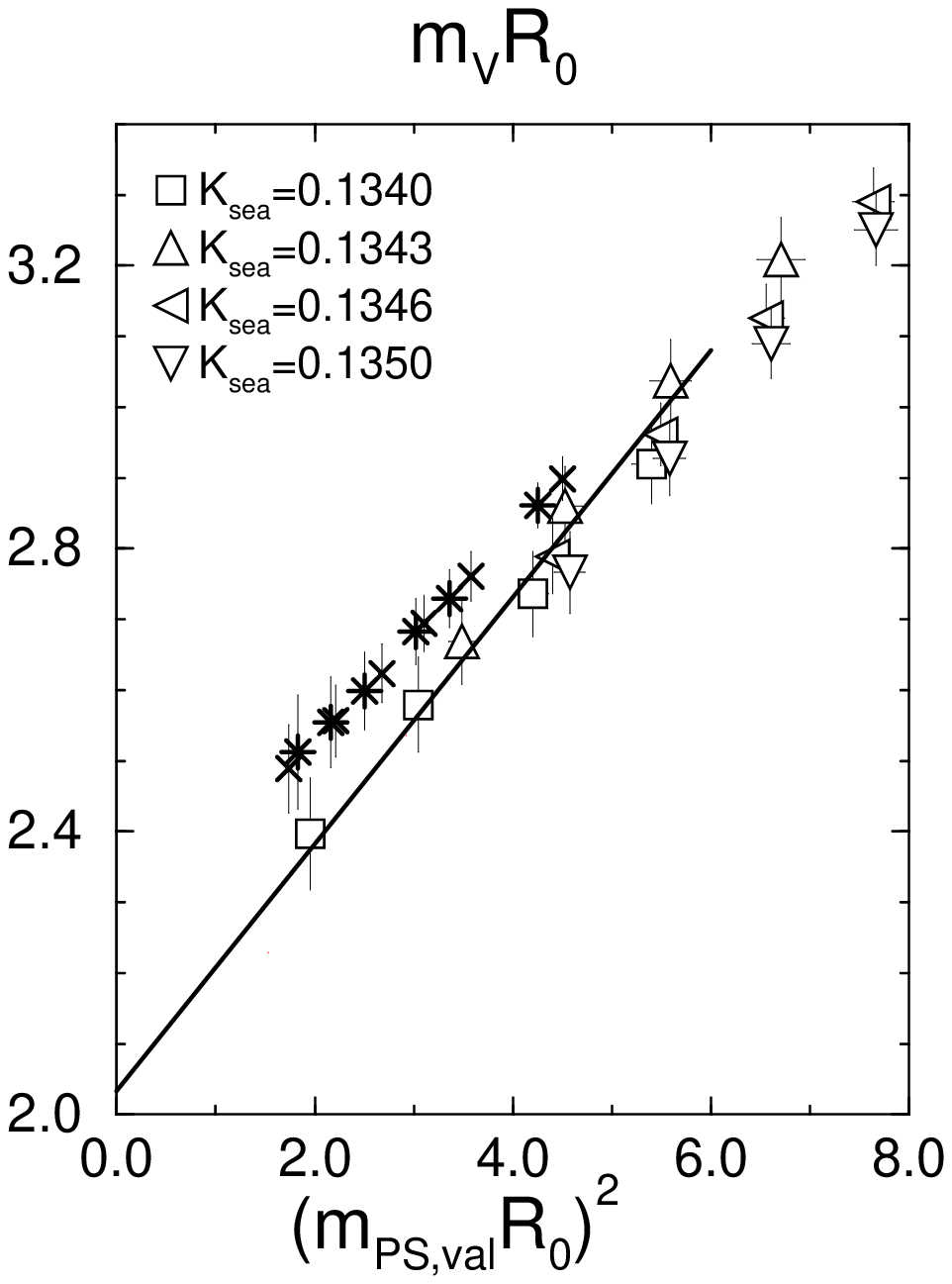} }
\vspace{-1cm}
\caption{(Left) $R_0\mV(s;v_1,v_2)$ as a function of
$(\mPSv R_0)^2$ for the PC(NP) action 
from the JLQCD collaboration\protect\cite{JLQCD_full},
together with the quenched data for the PC(NP) 
action\protect\cite{UKQCD_quench}.
Symbols are the same in Fig.~\protect\ref{fig:vps2}.
(Right) The same plot for the PC(NP) action
from the UKQCD collaboration\protect\cite{UKQCD_full_NP},
together with the fit line of the left figure.}
\label{fig:vps2_both}
\vspace{-0.5cm}
\end{figure}

To see if this is really a dynamical quark effect,
the same data normalized by $R_0$
are shown in the left  part of Fig.~\ref{fig:vps2_both},
together with quenched data at $\beta=6.0$ and 6.2 for the same 
action\cite{UKQCD_quench}. Note that
the scaling violation is negligible in the quenched case.

A curvature can not be clearly seen anymore, and
a linear decrease seems steeper than the quenched one.
Partly due to larger errors caused by that of $R_0$, however,
the fit by the form
\begin{eqnarray}
R_0\cdot \mV(s;v_1,v_2)&=& A + B_s\cdot R_0^2\cdot \mPS^2 \nn\\
&+& B_v \cdot R_0^2 \cdot \mPSv^2 
\end{eqnarray}
can not resolve a sea quark mass dependence in the data,
giving $B_s \simeq 0$. 
The fit with $B_s=0$, plotted in the figure, gives a reasonable value of
$\chi^2$, showing a steeper slope $B_v$, which
may be regarded as the dynamical quark effect.
Further precise calculations will be required, however, 
to resolve the sea quark mass dependence of the slope $B_s$
in this range of sea quark masses($\mPS/\mV\simeq 0.6-0.8$),
so that the dynamical quark effect is undoubtedly identified.

As mentioned before, the UKQCD collaboration has performed an unquenched
simulation varying both $\beta$ and $m_q a$ of PC(NP) action in order to
keep $R_0$ constant\cite{UKQCD_full_NP}.
As shown in the right part of Fig.~\ref{fig:vps2_both},
the result is consistent with the one
from the fixed $\beta$ simulation of the JLQCD 
collaboration\cite{JLQCD_full},
once both data are normalized by $R_0$.

\subsection{Chiral extrapolation}
The dynamical quark effect is expected to be enhanced toward
smaller sea quark masses reached by the chiral extrapolation,
which is briefly described in this subsection.

The light quark mass is determined by the experimental value of
$m_\pi/m_\rho$. In the present method, the ratio determines the corresponding
PS meson mass, $\mPSs a $ by solving the equation
\begin{equation}
\dfrac{m_\pi}{m_\rho} = \dfrac{\mPSs a}{A + (B_s+B_v)\mPS^2a^2 +\cdots}
\label{eq:light}
\end{equation}
where the denominator is obtained by setting $\mPSv^2 =\mPSs^2$ in 
eq.~(\ref{eq:vector}), and the lattice spacing $a$ is explicitly written.
Hereafter $\mPSs^2 a^2$ is fixed to the solution of this equation,
$\mPSs a =m_\pi a$.

There are several ways to determine the strange quark mass. 
The $K$-input uses the experimental value of $m_K/m_\pi$ to determine other 
strange mesons, 
while the $\phi$-input uses the experimental value of $m_\phi/m_\rho$

To compare results at different $a$, hadron masses are normalized by $r_0$,
the value of $R_0$ in the chiral limit.

\subsection{Meson spectra at dynamical light quark masses}
\begin{figure}[tbh]
\centerline{\epsfxsize=4cm \epsfbox{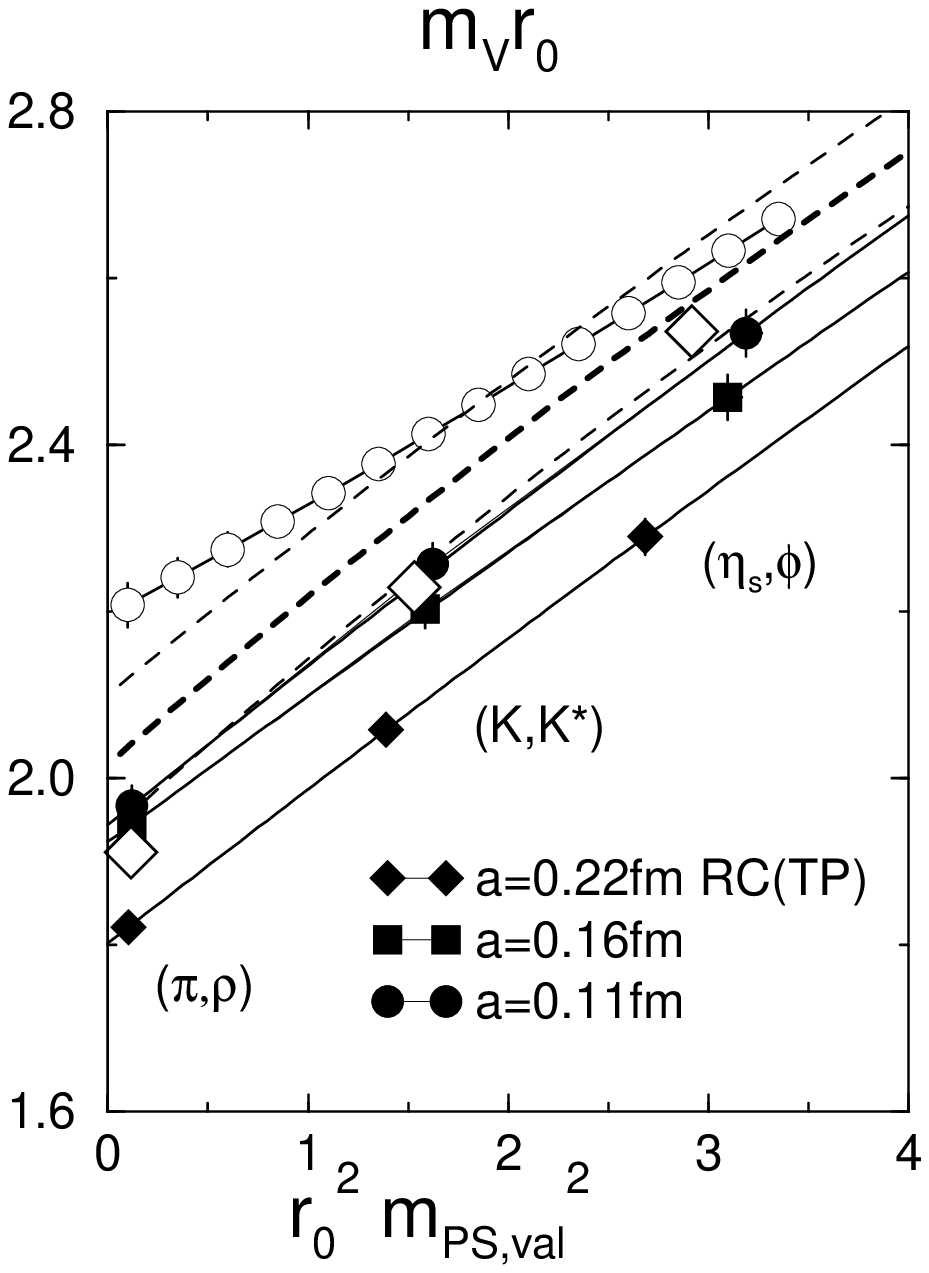}
            \epsfxsize=4cm \epsfbox{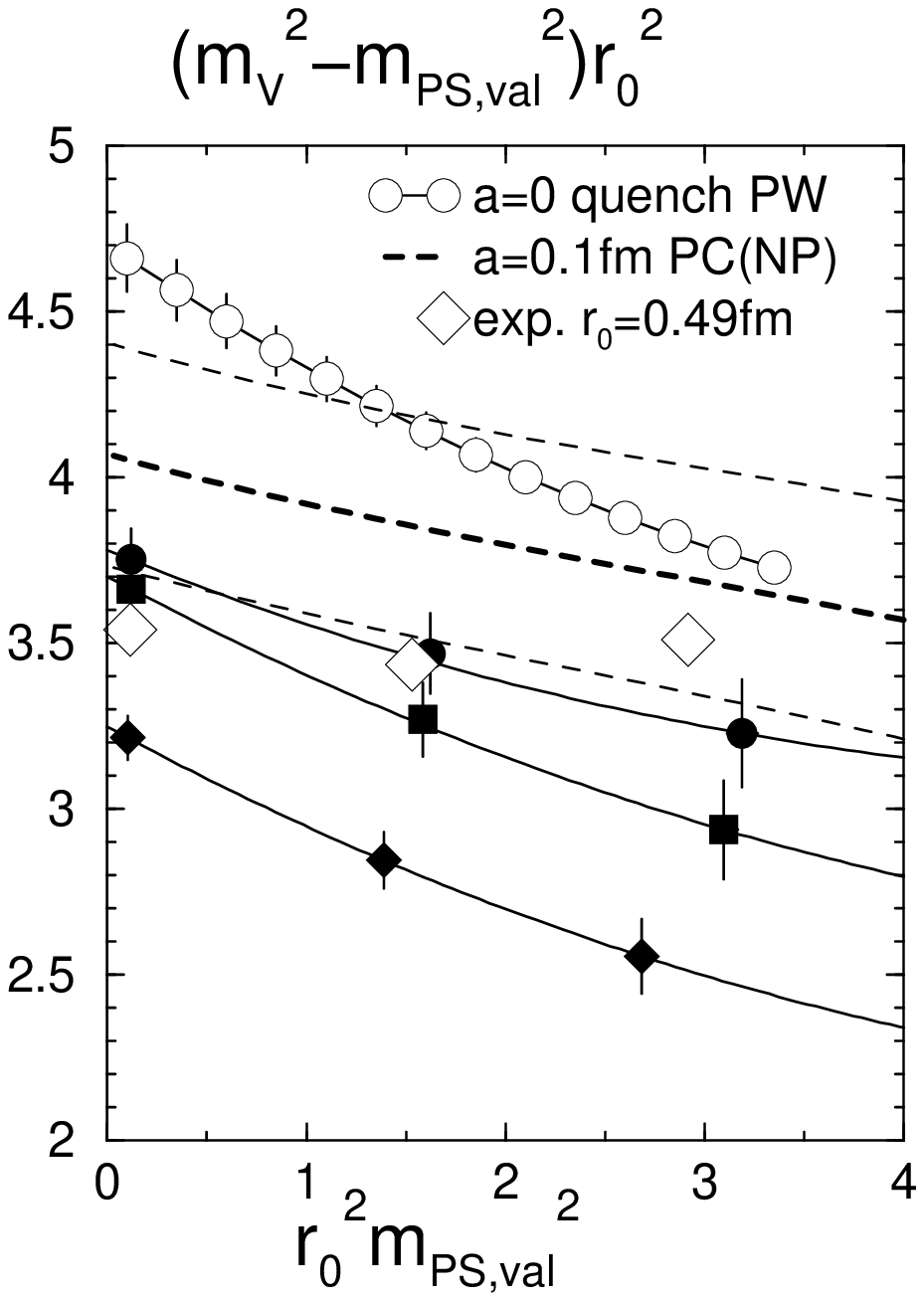}}
\vspace{-1cm}
\caption{(Left) $\mV r_0$(left) and  (Right) $(\mV^2-\mPSv^2) r_0^2$
as a function of $\mPSv^2 r_0^2 $ with $\mPSs^2 = m_\pi^2 $ fixed,
in unquenched QCD for RC(TP)\protect\cite{CPPACS_full}
and PC(NP)\protect\cite{JLQCD_full} actions,
together with the quenched result of the PW action\protect\cite{CPPACS_quench}.
}
\label{fig:vps_fixlight}
\vspace{-0.5cm}
\end{figure}

The vector meson mass normalized by $r_0$, $\mV r_0$ in unquenched QCD
is shown in the left part of Fig.\ref{fig:vps_fixlight}, 
as a function of $\mPSv^2 r_0^2$,
at $a=0.11\sim 0.22$ fm for the RC(TP) action\cite{CPPACS_full} and 
at $a=0.1$ for the PC(NP) action\cite{JLQCD_full},
together with the quenched result in the continuum limit\cite{CPPACS_quench}.
The lines in the figure are determined by eq.~(\ref{eq:vector})
with  $\mPSs^2=m_\pi^2$ fixed, and the points 
on the line marked by $(\pi,\rho)$, $(K,K^*)$ and $(\eta_s,\phi)$
are given by $\mPS/\mV =$ 0.1350/0.7684, 0.4977/0.8961 and
0.695/1.0194, respectively, where $m_{\eta_s}$=0.695 GeV is
used for an unphysical $\eta_s$.
The error of the line, not shown in the figure for the RC(TP) action, is 
roughly equal to the error of the points, where the error of $r_0$ is not 
included.
The error band for the PC(NP) action includes this source of error.

The slope of unquenched data increases as $a$ decreases, so that the slope
at $a=0.11$ fm of RC(TP) action is clearly steeper than the slope of the 
quenched result in the continuum limit, and closer to experiment.
This is the manifestation of dynamical quark effect on hadron spectra
we are looking for.
The chiral extrapolation in the sea quark mass, 
as well as control of scaling violation,
is necessary to see the effect clearly.
The line for the PC(NP) at $a=0.1$ fm, which lies a little above, seems 
consistent with the RC(TP) if errors from the remaining scaling violation and
$r_0$ are taken into account.

The result of $\mV$ can be converted into the hyper-fine splitting
of mesons, and is shown in the right part of Fig.~\ref{fig:vps_fixlight}, where
$(\mV^2-\mPSv^2) r_0^2$ is plotted as a function of $\mPSv^2r_0^2$
with $\mPSs^2 = m_\pi^2 $ fixed.
The slope of unquenched data decreases with $a$, and is very 
different from the quenched one in the continuum limit.
The dynamical quark effect is most noticeable for the hyper-fine 
splitting normalized by $r_0$.
\begin{figure}[tbh]
\centerline{\epsfxsize=7.2cm \epsfbox{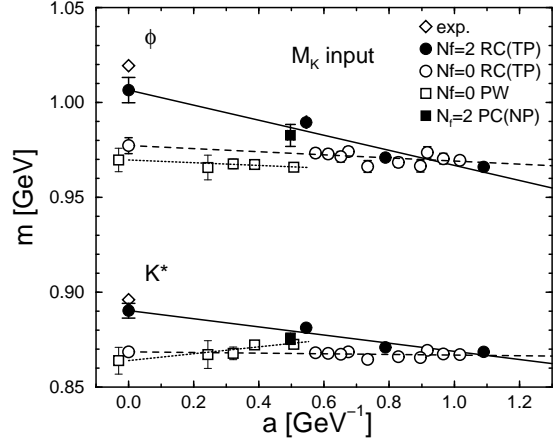}}
\vspace{-1cm}
\caption{$m_\phi$ and $m_{K^*}$(GeV) as a function of $a$ (GeV)
in $N_f =2$ QCD for RC(TP)\protect\cite{CPPACS_full}
and PC(NP)\protect\cite{JLQCD_full} actions and in quenched QCD for 
PW\protect\cite{CPPACS_quench} and RC(TP)\cite{CPPACS_full}
actions. The scale is set by $m_\rho$.}
\label{fig:meson-a}
\vspace{-0.5cm}
\end{figure}

Let me remark that a steeper slope of the hyper-fine splitting in quenched QCD
means a smaller hyper-fine splitting of strange mesons than the
experimental value, if
the splitting between $m_\rho$ and $m_\pi$ is fixed to the experimental value.

\subsection{Continuum limit of strange meson masses}
In Fig.~\ref{fig:meson-a}, the scaling behavior of $\phi$ and $K^*$ 
meson masses
from $K$-input, converted into GeV unit using $\rho$ meson mass,
are plotted in $N_f=2$ unquenched QCD for 
RC(TP)\cite{CPPACS_full} and PC(NP)\cite{JLQCD_full} actions,
together with the quenched result for PW\cite{CPPACS_quench} 
and RC(TP)\cite{CPPACS_full} actions.
Lines represent linear continuum extrapolations for the $N_f=2$ RC(TP),
quenched PW and RC(TP) results.

A large discrepancy between the experimental value
and the quenched result in the continuum limit 
is much reduced in $N_f$=2 unquenched QCD.
It will be interesting to see if
the inclusion of the dynamical strange quark
removes the remaining small difference.

\subsection{Remark on the continuum extrapolation}
Here a remark is given on the scaling behavior of 
perturbatively improved theories such as RC(TP), extensively employed  in 
unquenched QCD simulations.
Suppose that a renormalized observable $X(a)$ at non-zero lattice spacing $a$
is expanded in the bare coupling constant $g_0^2$ as
\begin{eqnarray}
X(a)&=&X_0 +g_0^2\left[ x_1^0 + a(x_1^1+y_1^1\log a)
\right] \nn  \\
& & + g_0^4\nonumber \left[x_2^0+y_2^0\log a + a(x_2^1 + y_2^1\log a 
\right. \nn \\
& & \left. + z_2^1\log^2 a) \right]+O(a^2,g_0^6)
\end{eqnarray}
Note that the $g_0^2 \log a$ term is absent because of the renormalizability.
The tree level $O(a)$ improvement leads to $y_1^1=0$ and $z_2^1=0$, while
the 1-loop $O(a)$ improvement gives $x_1^1=0$ and $y_2^1=0$ in addition.
In terms of the renormalized coupling at scale $\mu$ defined as
\[
g_0^2 = g_R^2 -\dfrac{y_2^0}{x_1^0}g_R^4\log (\mu a)
=\dfrac{g_R^2}{1-\dfrac{y_2^0}{x_1^0}\log (\mu a)}+\cdots ,
\]
one obtains
\begin{eqnarray}
X(a) &=& X(0) +g_R^2 a\left[ x_1^1+y_1^1\log a\right] 
\nn \\
&+ & g_R^4 a \left[x_2^1+y_2^1\log a + z_2^1\log^2 a \right] +\cdots\nn \\
X(0) &=& X_0 + g_R^2 x_1^0 + g_R^4 x_2^0 + O(g_R^6) \nn
\end{eqnarray}

The leading terms of scaling violation are
$g_R^2 a$ and $g_R^2 a\log a$ for unimproved theories such as the 
Wilson fermion,
$g_R^2 a$ and $g_R^4 a \log a$ for tree-level or tadpole $O(a)$ improved 
theories, or $g_R^4 a$ and $g_R^6 a \log a$ for 1-loop $O(a)$ improved 
theories.
This concludes that a linear continuum extrapolation employed for data in
the RC(TP) action\cite{CPPACS_full} is justified
if the $g_R^2 a$ term dominates
the $g_R^4 a\log a $ term and
$O(a^2)$ terms can be neglected.

Note that terms such as $(g_0^2)^n a \sim \dfrac{a}{(\log a)^n}$
never appear in the scaling violation,
since the physical scale $\mu$ is independent on $a$.

\subsection{Dynamical quark effect on baryon masses}
Finally the dynamical quark effect on baryon masses is briefly discussed.
The chiral extrapolation of baryon masses is made
in terms of the PS meson mass, by the formula explicitly given as
\begin{eqnarray}
&& m_\Sigma (s;v_1,v_2,v_2) = A + B_s \mu_s
+ 2 F\mu_{v_2} \nn \\
&+& (F-D)\mu_{v_1} + C_s \mu_s^2 + (E + E_O) \mu_{v_1}^2 \nn\\
&+& (E-E_O) \mu_{v_2}^2 + (G+G_O) \mu_s \mu_{v_1}  \nn \\
&+ & (G-G_O) \mu_s \mu_{v_2} + C_{v_{12}} \mu_{v_1}\mu_{v_2}
\end{eqnarray}
for $\Sigma$-like octet baryons, where
$\mu_s = \mPSs^2$ and $\mu_{v_i} = \mPS^2(s;v_i,v_i)$.
Similar formula are given for the $\Lambda$-like octet baryon
mass, $m_\Lambda (s;v_1,v_2,v_2)$, and simpler ones
for the decuplet baryon mass $m_D (s;v_1,v_2,v_3)$\cite{CPPACS_full}.
\begin{figure}[tbh]
\centerline{\epsfxsize=7.5cm \epsfbox{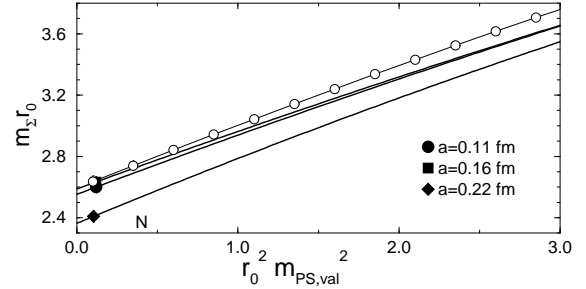}}
\vspace{-1cm}
\caption{$\Sigma$-like octet baryon masses, normalized by $r_0$,
as a function of $r_0^2\ \mPSv^2$ 
for the unquenched RC(TP) action\protect\cite{CPPACS_full} and the quenched PW 
action\protect\cite{CPPACS_quench}.}
\label{fig:nps2_fixlight}
\vspace{-0.5cm}
\end{figure}

In Fig.~\ref{fig:nps2_fixlight} the $\Sigma$-like octet baryon masses
are plotted as a function of the averaged PS meson mass squared, 
$\mPSv^2 = (\mu_{v_1}+2\mu_{v_2})/3$, for the RC(TP) 
action\cite{CPPACS_full}.

To draw lines in the figure, the dynamical quark mass is kept fixed to the
light quark mass, $\mu_s = m_\pi^2$.
The solid line corresponds to $r_0 m_\Sigma (s; v,v,v)$ as a function of
$r_0^2 \mu_v $, 
while the solid symbols
represent $N$ at the point of $v=s$.
The error of each line, which does not include the error of $r_0$,
is roughly equal to the size of the symbol on 
the line.

The scaling violation is small at $ a \le 0.16$ fm and
the slope of lines in unquenched QCD becomes flatter than
that in quenched QCD, as $a$ decreases. 
This may be a dynamical quark effect. 
One must, however, be careful to conclude this,
since the physical lattice size in the simulations are different, 
$\simeq 2.5$ fm
in unquenched simulations and 3 fm in quenched ones, so that finite size
effects could make such a difference. It is therefore interesting to obtain the
corresponding plot in quenched QCD with the lattice size of 2.5 fm 
in the continuum limit, to identify the dynamical quark effect on the slope.
Moreover, even though some finite size effect is expected to exist 
on baryon spectra, the small scaling violation 
enables us to predict the chiral behavior of the baryon masses 
in the continuum
limit with a fixed lattice size of 2.5 fm.

Similar results are obtained for the $\Lambda$-like octet baryon mass and
the decuplet baryon mass.
In particular almost no scaling violation is observed for
the decuplet baryon mass at $a \le 1.6$ fm.

\section{Dynamical quark effect on other quantities}
\subsection{String breaking}
\begin{figure}[tbh]
\centerline{\epsfxsize=6cm \epsfbox{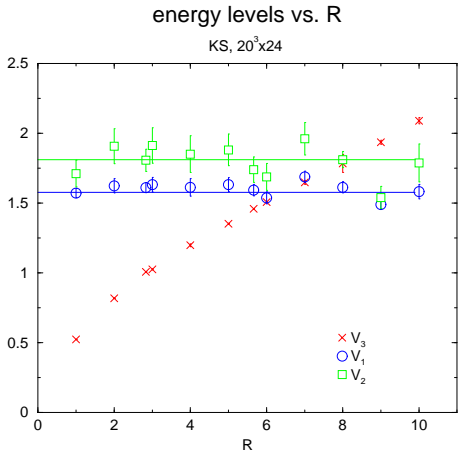}}
\vspace{-1cm}
\caption{Lowest 3 energy levels, corresponding two S-wave mesons,
S+P wave mesons and the string, as a function of the distance $R$.}
\label{fig:SB00}
\vspace{-0.5cm}
\end{figure}
In the past unquenched simulations, 
the expected behavior for string breaking that
the static quark potential $V(R)$ levels off at large enough $R$,
has not been observed\cite{SB_review,CPPACS_sb,SESAM_sb}: 
$V(R)$ increases linearly in $R$ beyond the point where it exceeds
the energy of the two mesons,
due to poor overlap of Wilson loops with the meson-meson state.

Several variational methods have been introduced to find better operators 
which have larger overlap with the meson-meson state,
and successfully applied to several models
such as an adjoint Wilson loop in pure gauge theories or gauge-Higgs models,
showing the expected behavior for ``string breaking''\cite{SB_review}. 

At lattice'99 the MILC collaboration has reported a first evidence of 
string breaking in unquenched QCD simulations\cite{MILC_sb99},
and the follow-up of the result is presented at this 
conference\cite{MILC_sb00}.
They employ the $N_f =2$ KS fermion for dynamical quarks on
a $(3.3 {\rm fm})^3 \times (3.9 {\rm fm})$ lattice at $a \simeq 0.16$ fm 
and $\mPS/\mV = 0.36$. 
The energy obtained from meson-meson as well as meson-string correlators
starts to level off at $R\simeq 1$ fm, where the energy is roughly equal to
twice the mass of the heavy-light meson\cite{MILC_sb99}.
The lowest three eigen-values, obtained by
the diagonalization of correlators including the string-string one,
are shown in Fig.~\ref{fig:SB00}\cite{MILC_sb00}, which supports  
the result in ref.\cite{MILC_sb99}.
The mixing between the string state and the meson-meson state, however,
is found to be too weak to show level crossing\cite{MILC_sb00}.

Let me give one comment on the definition of string breaking.
String breaking should be a
phenomenon that the energy for one $R$-independent operator 
linearly increases at short distance and levels off
at long distance. According to this criterion the behavior in 
ref.~\cite{MILC_sb99} is really string breaking.
If one uses different operators at different $R$ instead, 
the lowest energy can always be obtained by a diagonalization,
even in the absence of the transition from the string state to the
meson-meson state.

Apart from the string breaking, decays of hadrons, such as $\rho\to \pi\pi$,
give some indications of the dynamical quark effect, since such decays are
possible but largely suppressed in the quenched approximation.
At this conference, the MILC collaboration presents a result
which indicates the decay of $a_0(0^{++})\to \pi(0^{-+}) \eta(0^{-+})
$\cite{MILC_decay}.
Contrary to the case of string breaking, the $O^{++}$ operator employed
in the simulation seems to well overlap with both $a_0$ and $\pi\eta$ 
states.

\subsection{Topological susceptibility}
From the flavor-singlet axial Ward-Takahashi(WT) identity in the continuum, 
\begin{eqnarray}
& &\langle (\partial^\mu A_\mu (x) + 2 m_q P(x) ) Q(y) \rangle \nn\\
&+& 2 N_f \langle Q(x) Q (y) \rangle = 0 ,
\end{eqnarray}
one obtains
\begin{eqnarray}
- 2 m_q\int d^4x\ \langle P(x) Q\rangle &=& 2 N_f \langle Q^2\rangle ,
\end{eqnarray}
where $Q=\int d^4x\ Q(x)$ is the topological charge with the topological 
charge density $Q(x) =\dfrac{g^2}{32\pi^2} \dfrac{1}{2}F_{\mu\nu}(x) 
\widetilde{F}^{\mu\nu}(x)$.
\begin{figure}[tbh]
\centerline{\epsfxsize=7.5cm \epsfbox{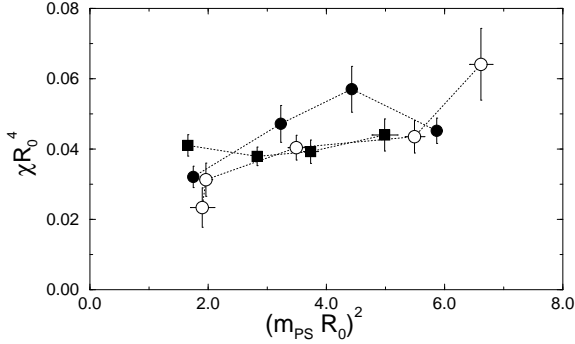}}
\vspace{-1cm}
\caption{$\chi R_0^4$ as a function of $\mPSs^2 R_0^2$
in $N_f=2$ unquenched QCD for the  PC(NP) 
action(open circles)\protect\cite{UKQCD_Q} at $a\simeq 0.1$ fm
and for the RC(TP) action(solid symbols) at
$a\simeq 0.11$ fm(circles)\protect\cite{CPPACS_Q} and  0.16 fm(squares)
\protect\cite{CPPACS_full}.
}
\label{fig:chi00}
\vspace{-0.5cm}
\end{figure}
Therefore 
\begin{eqnarray}
\chi &\equiv& \dfrac{\langle Q^2\rangle}{V} =-\dfrac{m_q}{N_f V}\int d^4 x
\langle P(x) Q \rangle \rightarrow 0
\end{eqnarray}
as $m_q\rightarrow 0$. The topological susceptibility $\chi$ must vanish 
in the chiral limit.
An investigation of this property on the lattice becomes a good check 
for lattice chiral symmetry in unquenched simulations.
The decrease of $\chi$ toward $m_q =0$ is driven purely by the fermion 
determinant.

In the past simulations, the expected decrease of $\chi$ has not been observed
for both KS\cite{PISA_Q} and Wilson-type\cite{CPPACS_full} quark actions.
The topological susceptibility normalized by $R_0$
does not show the quark mass dependence at all
and values are consistent with the quenched one\cite{CPPACS_full}.

This year  
the UKQCD collaboration\cite{UKQCD_Q} and CP-PACS
collaboration\cite{CPPACS_Q}
have obtained new results, which are shown in Fig.~\ref{fig:chi00}.
The decrease of $\chi$ toward the chiral limit
is observed at $a\simeq 0.1$ fm for both PC(NP) and RC(TP) actions, 
in contrast to the previous 
result at $a=0.16$ fm for the RC(TP) action, though it is still
far from verifying the flavor-singlet axial WT identity.
Theoretical considerations\cite{rossi} as well as further numerical 
analyses are needed to settle this issue completely.

\subsection{Flavor-singlet PS meson}
The fact that the flavor-singlet pseudo-scalar meson is not a 
Nambu-Goldstone boson(U(1) problem) must be demonstrated by lattice QCD
simulations. Numerically this has been a difficult task since 
one has to calculate the quark 2-loop diagram special for the 
flavor-singlet meson.
By the volume-sources method without gauge-fixing\cite{kuramashi} or
the U(1) random noise method\cite{kilcup_u1} the estimation of 
the flavor-singlet meson mass becomes possible in quenched QCD.
\begin{figure}[tbh]
\centerline{\epsfxsize=7.5cm \epsfbox{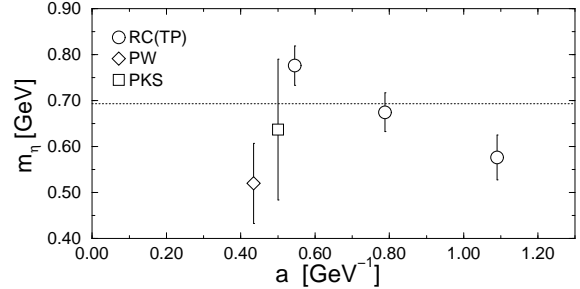}}
\vspace{-1cm}
\caption{$m_\eta$ in GeV as a function of $a$ in $N_f=2$ unquenched QCD
for RC(TP)\protect\cite{CPPACS_full}, PW\protect\cite{SESAM-TCL_eta}
and PKS\protect\cite{kilcup_u1}. The smearing source is employed for
the latter two. The horizontal line is the ``experimental'' value of
the flavor-singlet meson mass in $N_f=2$ QCD.}
\label{fig:eta}
\vspace{-0.5cm}
\end{figure}

The calculation of the singlet meson mass appears to be much
harder in unquenched QCD, where the individual meson propagators
should behave as
\begin{eqnarray}
\langle \eta (0) \eta (t) \rangle_{\mbox{2-loop}}
&\simeq& Z_\pi e^{-m_\pi t} - Z_\eta e^{-m_\eta t} \\
\langle \eta (0) \eta (t) \rangle_{\mbox{1-loop}}
&\simeq& Z_\pi e^{-m_\pi t} .
\end{eqnarray}
Here $\eta = \bar q \gamma_5 {\bf 1} q$ is the flavor-singlet PS meson 
operator,
$m_\eta$ is the corresponding mass 
and $m_\pi$ is the mass of the degenerate non-singlet PS meson.
Since the pole of the $\eta$ appears after the leading contribution of
$\pi$ is completely canceled between two diagrams, 
the signal becomes too noisy at large $t$ to extract $m_\eta$.

Therefore one is forced to extract $m_\eta$ at small $t$, and
one usually fits the ratio of propagators as\cite{kilcup_u1}
\[
R(t)=\dfrac{\langle \eta (0) \eta (t) \rangle_{\mbox{2-loop}}}
{\langle \eta (0) \eta (t) \rangle_{\mbox{1-loop}}}
\simeq 1 - \dfrac{ Z_\eta}{ Z_\pi} e^{-(m_\eta -m_\pi) t},
\]
hoping 
that a large part of contributions from excited states may be canceled out 
in the ratio.
The CP-PACS collaboration\cite{CPPACS_full} indeed found that
$m_\eta -m_\pi$ extracted from the ratio for $t_{min} = 2$
is consistent with the one for $t_{min} =3$, 4 within statistical
errors, where $t_{min}$ is the minimum value of $t$ for the fit .

Recently the SESAM-T$\chi$L collaboration has shown\cite{SESAM-TCL_eta} that
the plateau for $m_\eta$ as well as $m_\pi$ starts
at reasonably small $t$, employing the smearing source.
This opens a possibility for a reliable calculation of $m_\eta$
in very near future.

In the chiral limit one obtains a non-zero value of 
$m_\eta$\cite{kilcup_u1,CPPACS_full,SESAM-TCL_eta},
which is shown in Fig.~\ref{fig:eta} as a function of $a$, together
with the value of $m_\eta\simeq 0.7$ GeV for $N_f=2$ QCD
derived from the experimental value($N_f=3$)
by the Witten-Veneziano formula\cite{WV}.
Although results from different actions seems consistent within large
errors, 
the result from the RC(TP) action show
a systematic increase toward the continuum limit.
One has to check if  this large scaling violation
is caused by the contamination from excited states to the ground
state in the ratio $R(t)$.

A new calculation with the smearing source on configurations of 
the CP-PACS collaboration will make it possible to
extract $m_\eta$ reliably in the continuum limit
of $N_f=2$ QCD,
while the inclusion of the mixing to $\bar s s$ state\cite{UKQCD_eta} as
well as the dynamical strange quark will be necessary for the final result
in $N_f=3$ QCD.

\section{Summary}
We surprisingly find that hadron spectra from quenched QCD disagree in the 
continuum limit between KS and Wilson quark actions.
This discrepancy has to be resolved as soon as possible.
Personally I feel that some problems may exist in the identification of 
particles in KS quark actions, in particular of baryons. 
The restoration of the flavor breaking in hadron spectra might be a good check.
Hadron spectra from domain-wall and overlap quark actions\cite{vranas}
may also give an important hint for the solution.

Within Wilson-type quark actions,
the $N_f=2$ dynamical quark effect on hadron spectra, 
in particular in meson masses, has been demonstrated. 
Both chiral and continuum extrapolations are important to identify
the effect.
Results for baryons suggest that the scaling violation is small but
the lattice size has to be increased for reliable predictions in the continuum
limit.
Decreasing the dynamical quark mass will also be the next challenge in 
unquenched 
simulations.

The dynamical quark effect becomes manifest in some other quantities such as
the string breaking, the topological susceptibility and the flavor-singlet 
meson mass, even at finite lattice spacing.

New direction of investigations in unquenched QCD will be 
opened by  realistic $N_f=2+1$
unquenched simulation, which is the last step to numerically solve QCD.
Promising algorithm for $N_f=2+1$ QCD seems HMC for $N_f=2$ 
light quarks
and PHMC\cite{FJ} for the $N_f=1$ strange quark.
Some results will be expected at the first lattice conference in the next 
century.

\section*{Acknowledgements}
I would like to thank 
C. Allton, J.M. Carmona, A. Hart, A. Irving, K. Ishikawa,
D.B. Leinweber, J.W. Negele, H. Panagopoulos,
R. Parthasarathy, G. Rossi, S. Sharpe,
D. Toussaint, U. Wolff and T. Yoshie
for communicating and providing their results,
and, in particular, R. Burkhalter, C. DeTar, S. Gottlieb and T. Kaneko in 
addition for
making new data I have requested.
I also thank
R. Burkhalter, Y. Iwasaki, T. Kaneko, S. Hashimoto and A. Ukawa
for valuable comments on the manuscript.
This work is supported in part by the Grants-in-Aid of Ministry of 
Education(Nos. 12014202, 12640253).

\end{document}